# “Extraordinary” Phase Transition Revealed in a van der Waals Antiferromagnet

Xiaoyu Guo[1, Δ], Wenhao Liu[2, Δ], Jonathan Schwartz[3], Suk Hyun Sung[3], Dechen Zhang[1], Makoto Shimizu[4,*], Aswin L. N. Kondusamy[2], Lu Li[1], Kai Sun[1], Hui Deng[1], Harald O. Jeschke[5], Igor I. Mazin[6], Robert Hovden[3], Bing Lv[2, +] and Liuyan Zhao[1, +]

[1] *Department of Physics, University of Michigan, Ann Arbor, MI 48019, USA*

[2] *Department of Physics, the University of Texas at Dallas, Richardson, TX 75080, USA*

[3] *Department of Materials Science and Engineering, University of Michigan, Ann Arbor, MI 48109, USA*

[4] *Department of Physics, Okayama University, Okayama 700-8530, Japan*

[5] *Research Institute for Interdisciplinary Science, Okayama University, Okayama 700-8530, Japan*

[6] *Department of Physics and Astronomy, and Quantum Science and Engineering Center, George Mason University, Fairfax, VA 22030, USA*

[+] Corresponding to: blv@utdallas.edu, lyzhao@umich.edu

[*] Present address: *Department of Physics, Graduate School of Science, Kyoto University, Kyoto 606-8502, Japan*

[Δ] Authors contribute equally

**While the surface-bulk correspondence has been ubiquitously shown in topological phases, the relationship between surface and bulk in Landau-like phases is much less explored. Theoretical investigations since 1970s for semi-infinite systems have predicted the possibility of the surface order emerging at a higher temperature than the bulk, clearly illustrating a counterintuitive situation and greatly enriching phase transitions. But experimental realizations of this prediction remain missing. Here, we demonstrate the higher-temperature surface and lower-temperature bulk phase transitions in CrSBr, a van der Waals (vdW) layered antiferromagnet. We leverage the surface sensitivity of electric dipole second harmonic generation (SHG) to resolve surface magnetism, the bulk nature of electric quadrupole SHG to probe bulk spin correlations, and their interference to capture the two magnetic domain states. Our density functional theory calculations show the suppression of ferromagnetic-antiferromagnetic competition at the surface responsible for this enhanced surface magnetism. Our results not only show unexpected, richer phase transitions in vdW magnets, but also provide viable ways to enhance magnetism in their 2D form.**

Surfaces are always present in material systems of finite size. In systems featuring spontaneous symmetry breaking phase transitions, such as magnetism, the presence of surfaces has the potential to enrich their phase transitions[1-4]. Three distinct phase transitions were theoretically identified, namely "ordinary", "surface", and "extraordinary", as illustrated in Fig. 1a. The typical transition, where the surface and the bulk order simultaneously, is called "ordinary", and the one when the surface orders, but the bulk does not, is a "surface" transition. The transition establishing the bulk order, while the surface one is already present, is called "extraordinary". The point where three different phases meet is a "special point". In the "ordinary" case, the bulk order generates an effective field to induce a finite order at the surface, even though the surface interaction ($J_s$) is weaker than the bulk one ($J_b$), resulting in a single phase transition. Conversely, in the "surface" case, the surface order cannot provide a notable effective field deep in the bulk, and therefore, the bulk undergoes a separate "extraordinary" phase transition, leading to two phase transitions in the region when $J_s$ is stronger than $J_b$.

Separation of an ordinary phase transition into surface and extraordinary ones is highly uncommon, and requires that interactions responsible for the ordering are enhanced at the surface compared to the bulk. In three dimensional (3D) magnetic materials where the interaction between sites within the plane ($J_{||}$) is comparable to that for sites between the neighboring planes ($J_\perp$), the mean-field coupling at the surface ($J_s$), being the sum of all interactions, is expected to be smaller than the one inside the bulk ($J_b$) because of the loss of a $J_\perp$contribution from a neighboring layer (Fig. 1b). For such 3D materials, it is unlikely that any minor surface modifications could compensate for the missing $J_\perp$ that is of similar strength as $J_{||}$. On the other hand, in quasi-two-dimensional (2D) materials where $J_{||}$ is much larger than $J_\perp$ (Fig. 1c), it becomes possible that small changes in the surface layers could make up for the very weak missing $J_\perp$, or even push its mean field coupling strength beyond the bulk one to overcome the reduction due to stronger fluctuations at the surface. Therefore, quasi-2D materials, such as van der Waals (vdW) materials, are a potential material platform for realizing split surface-extraordinary phase transitions. Yet, the research on vdW and 2D materials in the past couple of decades hardly revealed any viable candidates for such splitting.

One major challenge for detecting surface and extraordinary phase transitions is the lack of experimental tools sensitive to phase transitions both at the surface and inside the bulk. The leading order electric dipole (ED) contribution to second harmonic generation (SHG) is known as an excellent probe for the broken spatial inversion symmetry and has been used extensively to investigate surface properties[5-9]. Very recently, the next order electric quadrupole (EQ) and magnetic dipole (MD) contributions to SHG have been successfully detected in many spatial-inversion-symmetric materials[10-13] and further emerged as an important tool for revealing centrosymmetric bulk phase transitions[14-19]. The combination of ED and EQ/MD contributions to SHG is a suitable tool for an experimental discovery of surface and extraordinary phase transitions.

The material candidate selected for this study is CrSBr, a vdW layered crystal with an orthorhombic point group ($mmm$). The structural primitive cell contains two edge-sharing distorted octahedra, with Cr at the center and S/Br at the vertices, forming an in-plane ($ab$-plane) orthorhombic network and stacking vertically along the out-of-plane ($c$-axis) direction[20, 21]. Bulk CrSBr exhibits four characteristic temperatures: $T^*$ = 185 K[22] and $T^{**}$ = 155 K[21-24] for two crossovers for the enhanced local dynamic spin correlations, $T_N$ = 132 K for the onset of bulk layered antiferromagnetic (AFM) order[20-27], and $T_F$ = 30 – 40K for the formation of a possible ferromagnetic (FM) state with debated origins[24, 26, 28]. The layered AFM features a FM spin alignment along the $b$-axis within each atomic layer and an AFM coupling between adjacent layers along the $c$-axis. The magnetic point group is $mmm1'$ for the bulk AFM order where the $c$-axis translational symmetry is present and $m'm2'$ for the surface order where the out-of-plane translational symmetry is absent. Interestingly, the onset of layered AFM occurs at different temperatures for CrSBr of different thicknesses: 138 K for six-layer CrSBr, 140 K for bilayer CrSBr and possibly 146 K for monolayer CrSBr[21]. This monotonic increase of the magnetic onset temperature with the decreasing thickness in CrSBr starkly contrasts with nearly all known vdW magnets, such as $CrI_3$[29, 30], $Cr_2Ge_2Te_6$[31], $Fe_3GeTe_2$[32], $NiPS_3$[33], $FePS_3$[34, 35], $MnPSe_3$[36], *etc.*, where the magnetic onset temperature in few-layer samples is either lower or equal to that of the bulk crystals.

Figure 2a summarizes the magnetic characteristic temperatures in CrSBr of different thicknesses that are probed by different experimental techniques. It can be seen that eliminating all neighboring layers, *i.e.*, transitioning from bulk to monolayer, leads to an increase in the critical temperature by possibly about 14 K, and that keeping only one neighboring layer, *i.e.*, going from bulk to bilayer, results in an increment of 8 K, close to half of the 14 K increment above. In addition, the Néel temperature of 138 K for six-layer CrSBr is comparable to the average temperature of two bilayers and four bulk layers, *i.e.*, $(2T_N^{bilayer} + 4T_N^{bulk})/6 =$135 K. Such considerations motivate us to investigate the surface of bulk CrSBr, closely resembling the bilayer case with only one neighboring layer, to search for the extraordinary and surface phase transition.

## Results

### STEM, TRANSPORT, AND SHG CHARATERIZATIONS OF 3D CrSBr CRYSTALS

vdW materials often suffer from atomic defects and stacking faults that potentially affect their electronic and magnetic properties[37-40]. To assess the crystallographic quality of our bulk CrSBr, we performed high-angle annular dark-field scanning transmission electron microscopy (HAADF-STEM) measurements in both plan-view and cross-section-view configurations (see Methods). Figure 2b shows the atomic structure of CrSBr within the *ab*-plane, with the Br/S column appearing to be the brightest, followed by a dimmer Cr column. Across multiple sites and samples of various thicknesses, atomic defects were rarely observed in the plan-view STEM images of CrSBr. Figure 2c displays the layered structure of CrSBr viewed in the *ac*-plane, with the vdW gaps showing up as the darker space between atomic layers. We further confirm that the overlying interlayer stacking is the sole preferred stacking geometry for CrSBr and barely any stacking faults were observed across multiple sites. This scarcity of atomic defects and stacking faults confirms the high crystalline quality of our CrSBr samples (see Methods).

The temperature-dependent heat capacity for single-crystalline CrSBr samples clearly reproduces the three temperature scales reported in literature, $T^*$ = 185 K, $T^{**}$ = 155 K, and $T_N$ = 132 K (Fig. 2d), whereas $T_F$ = 30 K is revealed by the magnetic susceptibility measurement (see Supplementary Section 1). Intriguingly, the temperature dependence of the SHG intensity from the same CrSBr batch exhibits a clear order-parameter-like upturn at 140 K (Fig. 2e), a new temperature scale for 3D CrSBr crystals, not detected by any probes for bulk properties, for example, heat capacity[21, 22], magnetic susceptibility[20, 22, 24, 26-28], neutron single crystal diffraction[23], and zero-field $\mu$SR[26]. We note that, first, unlike $T_N$ = 132 K for 3D CrSBr single crystals, neutron diffraction experiments on CrSBr powders that have a substantial surface to bulk ratio[26] showed $T_N$ = 140 K and, second, SHG revealed $T_N$ = 140 K for bilayer CrSBr whose composing layers miss the neighboring layer on one side[21]. The occurrence of the same critical temperature 140 K in bulk single crystals of this study, together with the surface sensitivity of ED SHG probe[5-9, 41, 42], indicates that this 140 K onset in bulk CrSBr crystals is likely the surface ordering temperature $T_S$ = 140 K, which is higher than the bulk Néel temperature, $T_N$ = 132 K, but lower than the crossover temperature, $T^{**}$ = 155 K (Fig. 2a).

## OBLIQUE INCIDENT RA SHG TRACKING PHASE TRANSITIONS IN 3D CrSBr CRYSTALS

To analyze further the magnetic phase transitions in bulk CrSBr crystals, we performed the rotation anisotropy (RA) measurements of SHG at an oblique incident angle $\theta$, to capture the symmetry evolution across the critical temperatures. In a SHG RA measurement (Fig. 3a), the intensity of reflected SHG light is recorded as a function of the azimuthal angle $\phi$ between the crystal axis $a$ and the light scattering plane in one of the four polarization channels, $P/S_{in}$-$P/S_{out}$, with $P/S_{in/out}$ standing for the incident/outgoing light polarization selected to be parallel/perpendicular to the light scattering plane. We start by showing SHG RA data taken at high temperatures ($T \geq T^*$), specifically at 185 K (Fig. 3b) and 295 K (see Supplementary Section 2), who are nearly identical in both the RA patterns and the SHG intensity in all four polarization channels. The four SHG RA polar plots in Fig. 3b are two-fold rotational symmetric about the $c$-axis ($C_{2c}$) and mirror symmetric with mirrors normal to the $a$- and $b$-axis ($m_a$ and $m_b$). They are well fitted by the EQ contribution to the SHG under the centrosymmetric point group $mmm$. We exclude the surface ED, bulk MD, and electric field-induced SH contributions as primary sources, even if present, for our SHG RA data at $T \geq T^*$ (Supplementary Section 2).

Upon cooling to low temperatures ($T < T_N$), we observe two, and only two distinct types of SHG RA data at 80 K through measurements across multiple thermal cycles and in different samples, as shown in Figs. 3c and 3d. Contrary to the RA patterns at 185 K, the SHG RA patterns at 80 K evidently break the $C_{2c}$ and $m_a$ symmetries but retain the $m_b$ symmetry. The comparison between Figs. 3c and 3d demonstrates that the two types of SHG RA data are related by either a $C_{2c}$ or a $m_a$ operation, which are the symmetries broken below $T_N$. Such a symmetry relationship between data in Figs. 3c and 3d confirms that these two types of SHG RA data correspond to two degenerate domain states that occur below $T_N$, corresponding to spins in the top layer being either parallel or anti-parallel along the $b$-axis and are labeled as Domain A and Domain B, respectively. A real-space survey of SHG RA across a CrSBr single crystal surface shows that the domain size extends up to 500 $\mu$m (see Supplementary Section 3), and a survey conducted over several thermal cycles demonstrated the random selection of domain states in individual thermal cycles (see Supplementary Section 3).

To model and fit the SHG RA data at low temperatures, we need to identify the SHG radiation sources and their corresponding point groups. Firstly, due to the absence of reported structural transitions for CrSBr within our temperature range of interest (80 K – 295 K)[26] , the EQ contribution to SHG based on the structural point group $mmm$ ($\chi_{ijkl}^{\mathrm{EQ,s}}$) should be present at all temperatures. Secondly, due to the centrosymmetric and time-invariant bulk layered AFM order that sets in below $T_N$ = 132 K, the EQ contribution to SHG from the magnetic point group $mmm1'$ ($\chi_{ijkl}^{\mathrm{EQ,bAFM}}$) should be considered at temperatures below $T_N$. We notice that the symmetry constraints are the same for the structural point group $mmm$ and the magnetic point group $mmm1'$, resulting in that $\chi_{ijkl}^{\mathrm{EQ,s}}$ is of the same form as $\chi_{ijkl}^{\mathrm{EQ,bAFM}}$.

Hence, from this point onward, we use $\chi_{ijkl}^{\mathrm{EQ}}$ to represent the combined contributions from the structure and the bulk AFM. Thirdly, the surface layered AFM breaks the spatial inversion and time reversal (TR) symmetries, and as a result, the ED contribution to SHG from the surface magnetic point group $m'm2'$ ($\chi_{ijk}^{\mathrm{ED}}$) should be included at low temperatures $T < T_{\mathrm{S}}$. Therefore, the RA SHG data at 80 K should be modeled by a coherent superposition of the EQ and the ED contributions (see Supplementary Section 4) and thus, is capable of probing the magnetic phase transition at the surface. The fitted results based on this model are depicted in Figs. 3c and 3d, and the interference between the EQ and ED contributions is illustrated for the $S_{\mathrm{in}}$-$S_{\mathrm{out}}$ polarization channel in Fig. 4.

Figure 4a confirms the sole presence of the EQ contribution to SHG at $T = 185$ K. More interesting is that Fig. 4b shows distinct consequences between the two domain states from the interference of the bulk EQ and the surface ED contributions: constructive interference in the top half and destructive interreference in the bottom half for Domain A, and the exact opposite way for Domain B. The bulk AFM order preserves all the symmetry operations in the structural point group and the TR operation, resulting in $\chi_{ijkl}^{\mathrm{EQ}}$(Domain A) = $\chi_{ijkl}^{\mathrm{EQ}}$ (Domain B). The surface AFM order however breaks $C_{2c}$, $m_{a}$, and TR symmetries that relate the two domain states, leading to $\chi_{ijk}^{\mathrm{ED}}$(Domain A) = $-\chi_{ijk}^{\mathrm{ED}}$(Domain B) (see Supplementary Section 4). This opposite sign relationship between the EQ and ED SHG susceptibilities for the two domain states explains the distinct interference behaviors observed in Fig. 4b.

Our next step is to track the magnetic phase transitions by performing careful temperature-dependent SHG RA measurements, paired with magnetic susceptibility measurements on the same CrSBr crystal. Figure 5a shows a color map of SHG intensity taken in the $S_{\mathrm{in}}$-$S_{\mathrm{out}}$ channel as functions of azimuthal angle $\phi$ and temperature $T$. A horizontal linecut at a fixed $\phi_0$ yields the temperature-dependent SHG intensity, such as the trace shown in Fig. 2e, whereas the vertical linecut at a selected $T$ gives the SHG RA pattern, such as the polar plots shown in Fig. 4. To better visualize the evolution of the SHG RA data as the temperature decreases, we present polar plots at four representative temperatures in the inset of Fig. 5a: $T = 185$ K (around $T$*), 146 K (between $T$** and $T_{\mathrm{S}}$), 138 K (between $T_{\mathrm{S}}$ and $T_{\mathrm{N}}$), and 80 K (below $T_{\mathrm{N}}$). A clear trend can be observed: the RA pattern first increases in the SHG intensity but retains the pattern shape of four even lobes until $T_{\mathrm{S}}$. It then exhibits two pairs of uneven lobes while further amplifying the intensity of the larger and reducing the intensity of the smaller pair below $T_{\mathrm{S}}$. As the temperature decreases below $T_{\mathrm{N}}$, a more pronounced contrast in the SHG intensity between the larger and smaller pairs of lobes is developed.

The SHG RA pattern at every temperature in Fig. 5a is fitted by the coherent superposition of the surface ED and the bulk EQ contributions to extract the temperature dependence of their sources. For the $S_{\mathrm{in}}$-$S_{\mathrm{out}}$ channel, the fitted results include two independent parameters for the surface ED source, $C_1^{\mathrm{ED}} = \chi_{xyy}^{\mathrm{ED}} + 2\chi_{yxy}^{\mathrm{ED}}$, and $C_2^{\mathrm{ED}} = \chi_{xxx}^{\mathrm{ED}}$, and another two for the bulk EQ one, $D_1^{\mathrm{EQ}} = \chi_{xxxx}^{\mathrm{EQ}} - 2\chi_{xxyy}^{\mathrm{EQ}} - \chi_{yxyx}^{\mathrm{EQ}}$, and $D_2^{\mathrm{EQ}} = \chi_{xyxy}^{\mathrm{EQ}} + 2\chi_{yyxx}^{\mathrm{EQ}} - \chi_{yyyy}^{\mathrm{EQ}}$. Because $\chi_{ijk}^{\mathrm{ED}}$ is variant and $\chi_{ijkl}^{\mathrm{EQ}}$ is invariant under the TR operation, we know that $\chi_{ijk}^{\mathrm{ED}}$ is proportional to the odd powers of the Néel vector (**N**) and $\chi_{ijkl}^{\mathrm{EQ}}$ scales with the even powers of **N**. Under the leading-order approximation, $\chi_{ijk}^{\mathrm{ED}} \propto \mathbf{N}$ and $\chi_{ijkl}^{\mathrm{EQ}} \propto constant + \mathbf{N}\cdot\mathbf{N}$, and as a result, $C_{1,2}^{\mathrm{ED}} \propto \mathbf{N}$ and $D_{1,2}^{\mathrm{EQ}} \propto constant + \mathbf{N}\cdot\mathbf{N}$. Figures 5b and 5c show the temperature dependence of $C_1^{\mathrm{ED}}$ and $D_1^{\mathrm{EQ}}$ (see $C_2^{\mathrm{ED}}$ and $D_2^{\mathrm{EQ}}$ in Supplementary Section 5), and Figure 5d displays the temperature dependence of bulk magnetic susceptibility of the same CrSBr crystal. The bulk magnetic susceptibility $\chi$ clearly shows a divergent behavior at $T_N$ = 132 K, as expected for a bulk CrSBr crystal[20-24, 26, 27]. The surface ED contribution $C_1^{ED}(T)$, which is proportional to $\mathbf{N}$, shows an order-parameter-like onset at $T_S$ = 140 K and then an observable kink at $T_N$ = 132 K. This observation confirms that the surface orders antiferromagnetically at a higher temperature than the bulk, providing definitive evidence for a surface phase transition in bulk CrSBr. The kink behavior at $T_N$ = 132 K reflects the impact of the bulk extraordinary phase transition on the surface order, which is consistent with the theoretical prediction of an $\beta > 1$ critical exponent for the surface order parameter at the extraordinary phase transition temperature[4, 43]. The EQ part $D_1^{EQ}(T)$, which scales with $\mathbf{N} \cdot \mathbf{N}$ after a constant offset from the structural contribution, initially experiences a steady but slow increase below $T^*$ = 185 K until $T^{**}$ = 155 K. Subsequently, it increases steeply across $T^{**}$, exhibits a notable peak at $T_S$ = 140 K, and ultimately a kink at $T_N$ = 132 K. Its capability to capture $T^{**}$ and $T_S$ stems from its sensitivity to the spin correlation via the term $\mathbf{N} \cdot \mathbf{N}$.

## FIRST-PRINCIPLE CALCULATIONS EXPLAINING SURFACE AND EXTRAORDINARY PHASE TRANSITIONS

To understand the increase in the magnetic onset temperature at the CrSBr surface despite the stronger Mermin-Wagner fluctuation expected at the surface, we refer to the Mermin-Wagner formula[44]: $T_N \approx \frac{T_{CW}}{A+\log(J_{||}/J')}$. Here, $T_{CW}$ denotes the mean-field transition temperature (in this case, simply the Curie-Weiss temperature) for monolayer CrSBr; $A$ is a constant of the order of 3-5; $J_{||}$ is the average characteristic intralayer exchange coupling; and $J'$ represents a properly-defined combination of the interlayer coupling $J_\perp$ and the Ising anisotropy $D$, which arises from both the single site anisotropy and the Ising exchange. In a previous study[45], $J'$ was estimated as $J' = D + J_\perp + \sqrt{(D+J_\perp)^2 - D^2}$. Note that in a single layer, the absence of $J_\perp$ can only lead to a decrease in $J'$ and consequently, $T_N$, assuming the same $T_{CW}$. In other words, our observed *increase* of $T_N$ at the surface with missing neighboring layers (i.e., $T_S$) must be attributed to the increase in $T_{CW}$. Note that $D$ is expected to be about the same in a single monolayer and in the bulk.

It is known that $T_{CW}$ depends on the intralayer exchange coupling and therefore on the lattice structure. To this effect, we performed careful first-principle density function theory (DFT) calculations (see Methods) to compute $T_{CW}$ based on the intralayer exchange coupling up to the 7th nearest neighbor (i.e., $J_{1-7}$) and for four structural configurations (S1–4, discussed below). The four strongest intralayer exchange couplings are found to be $J_1$, $J_2$, and $J_3$ that are FM, and $J_6$ that is AFM, as marked in Fig. 6a, whereas the remaining $J_4$, $J_5$, and $J_7$ are negligibly small (see Supplementary Materials Section 6). The four considered structures include bulk CrSBr (S1), rigid monolayer CrSBr that retains the atomic structure within the layer from the bulk (S2), fixed *ab* monolayer CrSBr that is derived from the intra-unit cell lattice relaxation while keeping the lattice constant same as the bulk (S3), and free monolayer CrSBr after the full lattice relaxation (S4). A holistic computation of $T_{CW}$ for the four structural cases reveals a consistent trend that is independent of the onsite Coulomb repulsion $U$: $T_{CW}$ increases from the bulk to the rigid monolayer, further enhances in the fixed *ab* monolayer, but decreases a bit in the free monolayer, as shown in Fig. 6b. This observed trend suggests two important factors for the enhancement of the magnetic onset temperature at the surface of bulk CrSBr: first, the absence of the neigboring layer, and second, the intra-unit cell lattice relaxation.

A close look into the evolution of the intralayer exchange coupling ($J_{1-7}$) across the four structures (S1 – 4) provides further insights into the two identified factors for the enhanced $T_{CW}$. The calculated $J_{1-3}$ show noticeable variations for S1 – 4, for a wide range of *U*, as shown in Figs. 6c-e, whereas $J_{4-7}$ remain unchanged across S1 – 4 structures (see Supplementary Materials Section 6). For the first factor, the absence of neighboring layers leads to a substantial increase in both $J_1$ and $J_2$, i.e., $\Delta_{S1\rightarrow S2} = \sim 1$ K shown in Figs. 6c and 6d. This increase in FM $J_1$ and $J_2$ by simply removing the neighboring layers is likely due to the suppression of a hopping path between the in-plane nearest ($J_1$) and second nearest ($J_2$) neighboring Cr sites going through neighboring layers that contributes an AFM intralayer exchange coupling to $J_1$ and $J_2$. For the second factor, the intra-unit cell lattice relaxation results in an increase in $J_2$ and $J_3$ by about 1 K and 2 K, respectively, i.e., $\Delta_{S2\rightarrow S3} = \sim 1$ K and $\sim 2$ K shown in Figs. 6d and 6e. The enhancement in $J_2$ is likely to arise from the increase of Cr-S-Cr angle between the two Cr sites within the unit cell (i.e., between the second nearest neighboring Cr sites), whereas the increase for $J_3$ mainly originates from the decrease of the Cr-S-Cr angle along the *b* axis (i.e., between the third nearest neighboring Cr sites). The strong dependence of the intralayer exchange coupling on the lattice structure has also been seen in strain-engineered CrSBr [46-48].

To summarize, we have successfully demonstrated the presence of surface and extraordinary phase transitions in a vdW AFM, bulk CrSBr, using the combination of bulk single-crystal EQ SHG and surface ED SHG. A clear temperature separation of 8 K is detected between the surface magnetism onset temperature, $T_S$ = 140 K, and the bulk Néel temperature, $T_N$ = 132 K. DFT calculations suggest two key factors for the increase of magnetic critical temperature at the CrSBr surface, namely, the absence of neighboring layer and the intra-unit cell lattice relaxation. Our results suggest multiple future research opportunities in vdW and 2D magnetism research. First, vdW magnets are a viable platform for realizing the parameter regime required for split surface and extraordinary phase transitions. In addition to CrSBr, immediate candidates include chromium chalcogenide halides[49, 50] and chromium oxyhalides[51] that have similar magnetic properties to CrSBr, and $VI_3$[52], which exhibits a similar thickness dependence of critical temperature; namely, the onset temperature is higher in the few-layer samples than in the bulk. Second, static strain[46-48, 53] and dynamic nonlinear phononics[54, 55] are promising ways to tune the magnetism or enhance the magnetic critical temperature of CrSBr, thanks to its extreme sensitivity of intralayer exchange coupling to the intra-unit cell atomic arrangement. It is quite likely that a large pool of vdW magnets exhibit a similar exchange coupling dependence on lattice structure as CrSBr and therefore can be candidates for strain and light engineering of magnetism. Third, moiré superlattice of CrSBr can be fundamentally distinct from that of $CrI_3$[56-60] and offer a new platform for exploring moiré magnetism. On the one hand, the intralayer exchange coupling in CrSBr is shown in this work to significantly depend on the presence of the neighboring layers, which can lead to periodical modulations of intralayer exchange coupling in twisted CrSBr superlattices. This is in contrast to twisted $CrI_3$ superlattices where only modulations in interlayer exchange coupling are considered. On the other hand, CrSBr has an orthorhombic crystal lattice with one-dimensional electronic properties[61, 62]. This highly anisotropic electronic property is in sharp contrast to the nearly isotropic electronic structure in $CrI_3$, and can offer unique moiré electronic and magnetic properties in twisted CrSBr.

## Figure Caption

**Figure 1 | vdW materials are a promising platform for hosting surface and extraordinary phase transitions. a**, A phase diagram illustrating ordinary, surface and extraordinary phase transitions and the special point. BD: bulk disordered, SD: surface disordered, SO: surface ordered, BO: bulk ordered. $J_s$, mean field surface interaction, $J_b$, mean field bulk interaction. **b** and **c**, Illustrations of the in-plane and the between-planes interactions, $J_{\parallel}$ and $J_{\perp}$, in **b,** 3D ionic crystals and **c**, quasi-2D van der Waals crystals.

**Figure 2 | STEM, heat capacity, and SHG characterizations of CrSBr bulk crystals. a,** Summary of magnetic phases and corresponding chracteristic temperatures in bulk and few-layer CrSBr from the literature. **b**, Plan- and **c**, side-view atomically resolved HAADF-STEM images of the CrSBr crystal, confirming the scarcity of atomic and stacking defects. **d**, Temperature dependent specific heat result showing the three reported characteristic temperatures, $T^*$ = 185 K, $T^{**}$ = 155 K, and $T_N$ = 132 K. **e,** Temperature dependent SHG intensity in the $S_{in}$-$S_{out}$ channel at the angle $\phi = 40°$ revealing an unreported onset at 140 K for bulk CrSBr. The red curve serves as a guide to the eyes.

**Figure 3 | SHG RA results revealing two degenerate magnetic domain states. a**, Schematic of the oblique incidence SHG RA measurement taken on a bulk CrSBr crystal. Red arrow: incident fundamental light, blue arrow: outgoing SHG light, gray arrows: light polarizations. **b**–**d**, SHG RA polar plots in four channels (P/$S_{in}$-P/$S_{out}$) at **b**, 185 K, **c**, 80 K from domain A and **d**, 80 K from domain B. Experiment data (circles) are fitted by functional forms simulated based on group theory analysis (solid curves). Numbers at the corners indicate the scales of the polar plots, with 1.0 corresponding to 1 fW.

**Figure 4 | Interference between bulk EQ and surface ED leading to distinct SHG RA patterns for the two domain states. a**, Left: schematic of the layered crystal structure at 185 K. Right: SHG RA pattern in the $S_{in}$-$S_{out}$ channel with only the EQ contribution. **b**, Left: schematic of the layered crystal structure overlaid with the spin texture in domain states A and B, related by the time-reversal operation ($TR$), two-fold rotation along the $c$-axis ($C_{2c}$) and mirror operation perpendicular to the $a$-axis ($m_a$). Right: SHG RA patterns in the $S_{in}$-$S_{out}$ channel, resulting from the interference between the bulk EQ and the surface ED contributions. The colored shaded areas of the SHG RA patterns indicate a $\pi$ phase shift of the SHG electric field from the white shaded areas. Stripped shaded areas indicate destructive interference. Numbers at the corners indicate the scales of the polar plots, with 1.0 corresponding to 1 fW.

**Figure 5 | Temperature-dependent SHG RA revealing the surface and the extraordinary phase transitions. a**, Lower: contour plot of the SHG RA in the $S_{in}$-$S_{out}$ channel as a function of temperature. Upper: SHG RA polar plots in the $S_{in}$-$S_{out}$ channel at four selected temperatures. **b**, $C_1^{ED}$ and **c**, $D_1^{EQ}$ as a function of temperature. Grey curves serve as guides to the eyes. **d**, Magnetic susceptibility as a function of temperature measured under 1000 Oe magnetic field along the $b$-axis. The regions of paramagnetism (PM), surface antiferromagnetism (s-AFM) and bulk antiferromagnetism (AFM) are shaded in different colors, with their characteristic temperatures marked.

**Figure 6 | DFT calculations explaining the origin of a higher transition temperature at the surface. a**, exchange pathways for $J_1$, $J_2$, $J_3$ and $J_6$, overlaid on the CrSBr crystal structure. **b**–**e**, $U$-dependence of **b,** $T_{CW}$ **c,** $J_1$ **d,** $J_2$, and **e,** $J_3$ for S1: bulk CrSBr (red), S2: rigid monolayer (orange), S3: fixed $ab$ monolayer (blue) and S4: free monolayer (green). $\Delta_{S1\rightarrow S2}$: change in $T_{CW}$ and corresponding $J$ from bulk to rigid monolayer. $\Delta_{S2\rightarrow S3}$: change in $T_{CW}$ and corresponding $J$ from rigid monolayer to fixed $ab$ monolayer.

**Acknowledgments:** L.Z., H.D., and K.S. acknowledge the support by the Office of Navy Research grant no. N00014-21-1-2770 and the Gordon and Betty Moore Foundation grant no. N031710. L.Z. also acknowledges the support by AFOSR YIP grant no. FA9550-21-1-0065 and Alfred P. Sloan foundation. B.L. acknowledges the support by US Air Force Office of Scientific Research Grant No. FA9550-19-1-0037, National Science Foundation (NSF)- DMREF- 1921581 and Office of Naval Research (ONR) grant no. N00014-23-1-2020. R.H. acknowledges the financial support of the W.M. Keck Foundation. This work made use of the Michigan Center for Materials Characterization (MC2). I.I.M. was supported by the U.S. Department of Energy through the grant No. DE-SC0021089. M.S. acknowledges the support by JSPS KAKENHI grant No. 22H01181. L.L. acknowledges the support by the National Science Foundation under Award No. DMR-1707620, by the Department of Energy under Award No. DE-SC0020184 (magnetization measurements).

**Author contributions:** X.G. and L.Z. conceived the project. X.G. performed the SHG measurements and data analysis under the supervision of L.Z., H.D., and K.S.. W.L. and B.L. provided the CrSBr single crystals. J.S., S.H.S. and R.H. performed the HAADF-STEM measurements. W.L., D.Z., L.L. and B.L. performed the heat capacity and magnetic susceptibility measurements. M.S., H.O.J and I.I.M performed and interpreted the DFT calculations. X.G. and L.Z. wrote the manuscript. All authors discussed the results.

**Competing Interests:** The authors declare no competing financial interests.

**Data Availability:** All data that support the finding of this work are included as main text and supplementary figures. Raw data and other data of this study are available from the corresponding author upon a reasonable request.

## Methods

**Crystal growth:** CrSBr single crystal is naturally grown using direct solid-vapor method through a box furnace. Cr powder (Alfa Aesar, 99.97%) and S powder (Alfa Aesar,99.5%) are accurately weighted inside the Ar-glovebox with total oxygen and moisture level less than 1 ppm. To facilitate the loading of bromide, the bromide liquid (99.8%) is initially solidified with the assist of liquid nitrogen. Cr powders, S powders and solid $Br_2$ are loaded into a clean quartz ampoule with the mole ratio of 1: 1.1: 1.2. Subsequently, the ampoule was sealed under vacuum using liquid nitrogen trap. We found out the extra amount of the S and Br created positive vapor pressure which effectively reduces the defects in the grown crystals and meanwhile promotes the larger size growth of single crystals. The quartz ampoule was heated up to 930°C very slowly, stay at this temperature for 20 hours, and followed by slow cooling down to 750°C (1°/hour). The assembly is then quenched down to room temperature. Large size CrSBr will grow naturally at the bottom of the ampoule. A small amount of $CrBr_3$ is also found at the top of the quartz ampoule and can be easily separated from the CrSBr crystals.

**Scanning tunneling electron microscopy (STEM):**

Plan-view specimens were prepared by exfoliating bulk CrSBr flakes on to polydimethylsiloxane (PDMS) gel stamps, which was transferred onto Norcada SiN TEM window grids with 2 μm holes. Cross-sectional specimens were prepared using the standard focus ion beam (FIB) lift-out method on Thermo Fisher Nova 200. HAADF-STEM was performed on JEOL 3100R05 (300 keV, 22 mrad) and Thermo Fisher Spectra 300 (300 keV, 21.4mrad) for plan-view and cross-section-view.

**Second harmonic generation:** The incident ultrafast light source was of 50 fs pulse duration and 200 kHz repetition rate with a center wavelength 800 nm. It was focused down to a 15 μm diameter spot on the sample with an oblique incidence angle $\theta = 11.2°$ and a power of 850 μW. The polarizations of the incident and reflected light could be selected to be either parallel or perpendicular to each other, with the azimuthal angle $\phi$ changing correspondingly. The intensity of the reflected SHG signal with 400 nm wavelength was detected by a charge-coupled device.

**First-principle calculations:** The structure relaxation of CrSBr bulk and monolayers was performed using the Vienna ab initio simulation package (VASP)[63-65]. The projector augmented wave (PAW) potentials with the generalized gradient approximation (GGA) exchange-correlation potential in the Perdew-Burke-Ernzerhof variant (PBE)[66] were used. For the bulk relaxation, we use a Γ-centered $6 \times 6 \times 4$ $k$ mesh, for monolayers a $7 \times 7 \times 1$ $k$ mesh and an energy cutoff of 900 eV. The convergence criterion is that all forces are smaller than 3 mV/Å. Then, we performed all electron density functional theory calculations using the full potential local orbit (FPLO) code[67]. **Energy mapping:** We use DFT energy mapping[68, 69] to determine the Heisenberg Hamiltonian parameters. For this purpose, we use a specially prepared 8-fold supercell of CrSBr with 16 symmetry inequivalent Cr sites. This allows us to determine the first seven in-plane exchange interactions $J_1$ to $J_7$. We use the GGA+$U$ exchange correlation functional with fully-localized limit double counting scheme, for eight different values of $U$ and $J_H$ = 0.72 eV fixed following Ref[70]. They were then fitted to the Heisenberg Hamiltonian of the form

$$H = \sum_{i<j} J_{ij} \boldsymbol{S}_i \cdot \boldsymbol{S}_j$$

with $Cr^{3+}$ spin operators $\boldsymbol{S}_i$ = 3/2. The Curie-Weiss temperature for this Hamiltonian is given by

$$T_{CW} = -\frac{1}{3} S(S+1)(2J_1 + 4J_2 + 2J_3 + 4J_4 + 4J_5 + 2J_6 + 4J_7)$$

where $S$ = 3/2. Figure 6a was generated using VESTA software[71].

# Figure 1

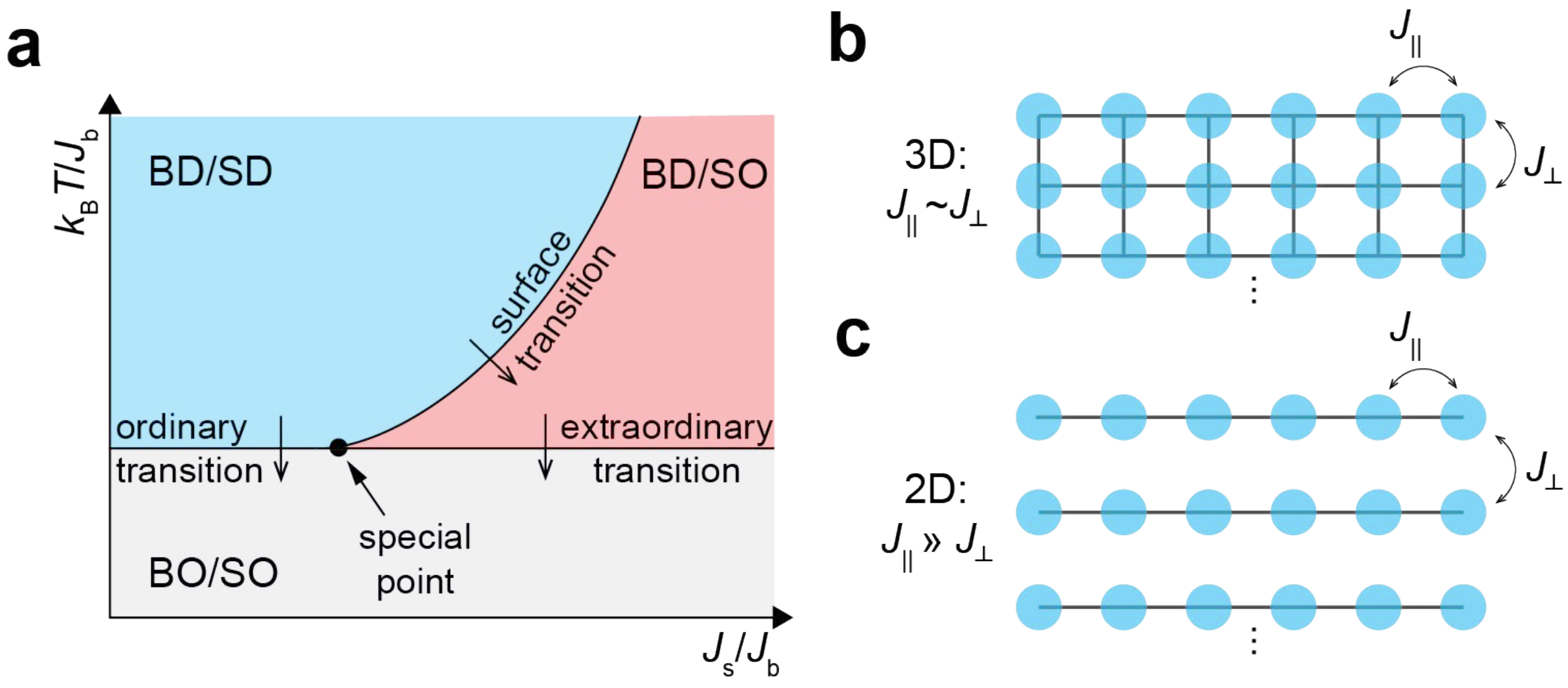

# Figure 2

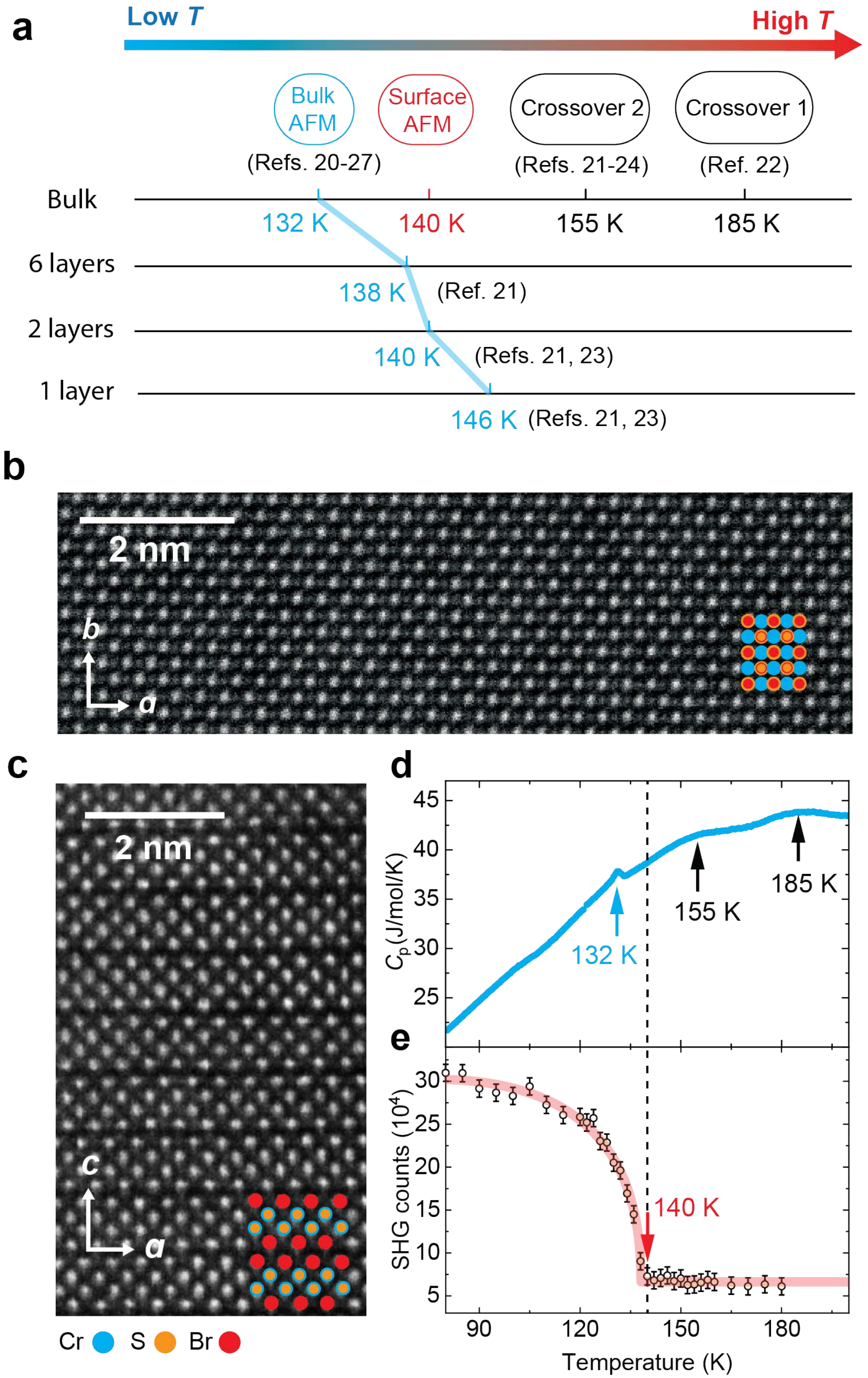

# Figure 3

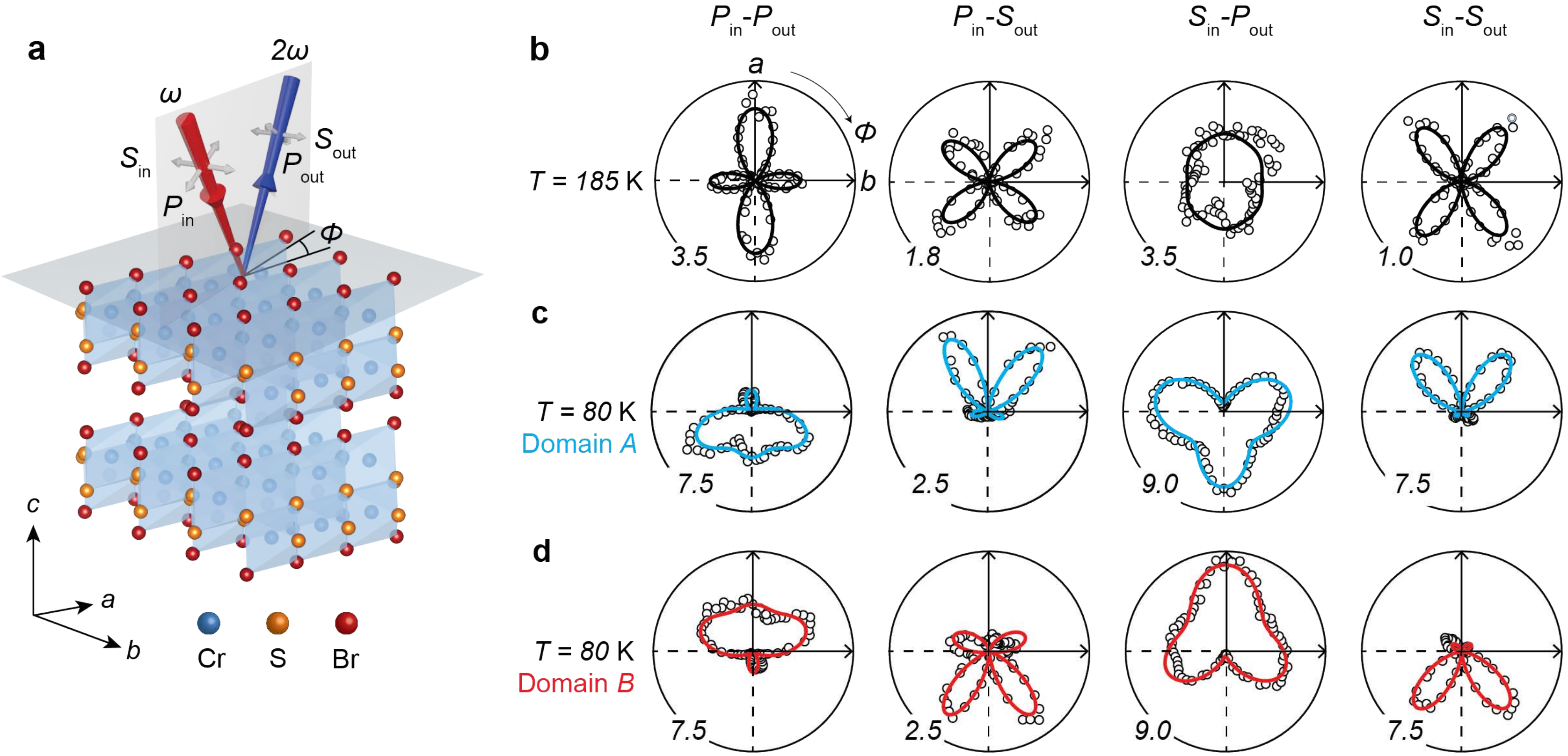

## Figure 4

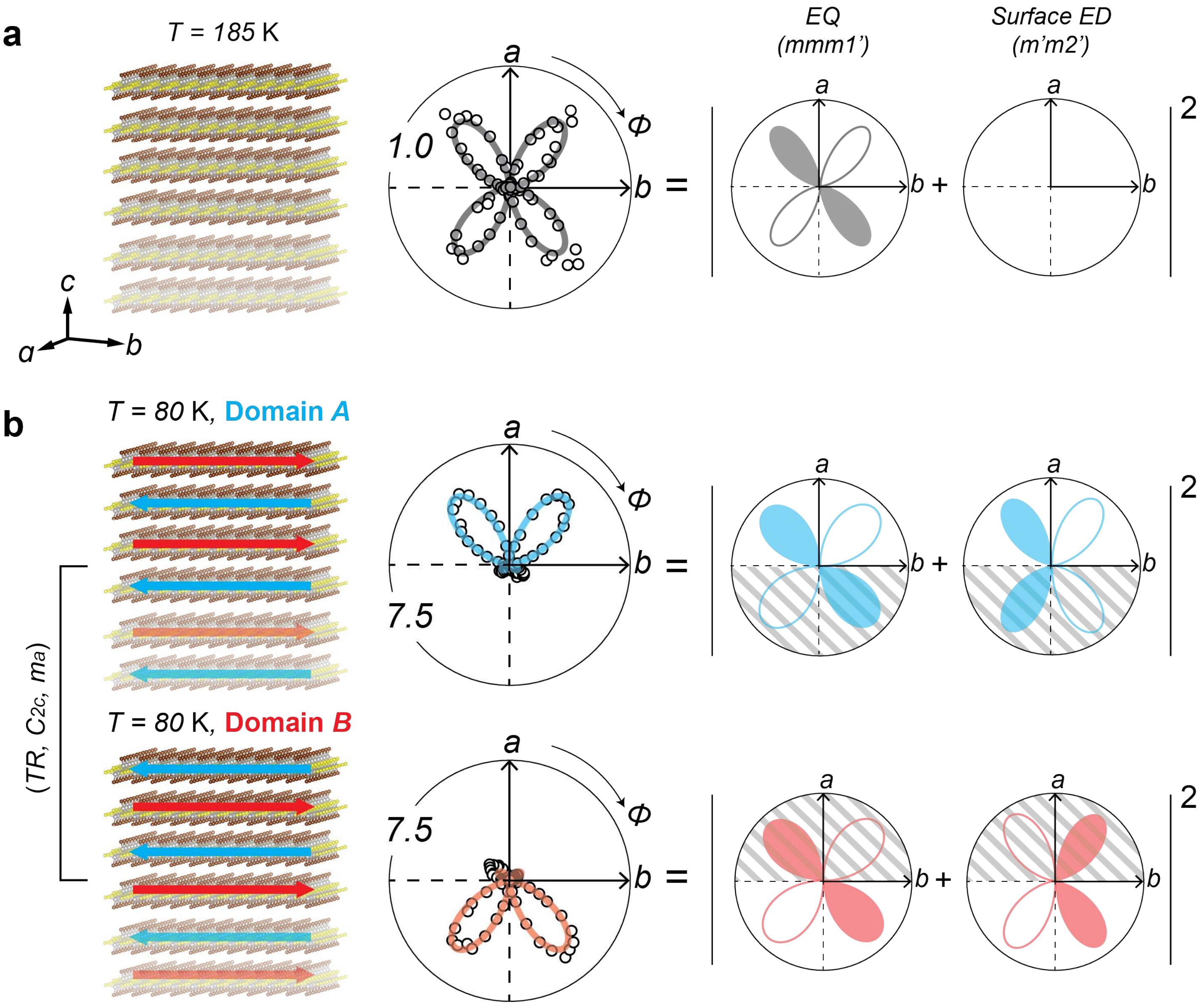

# Figure 5

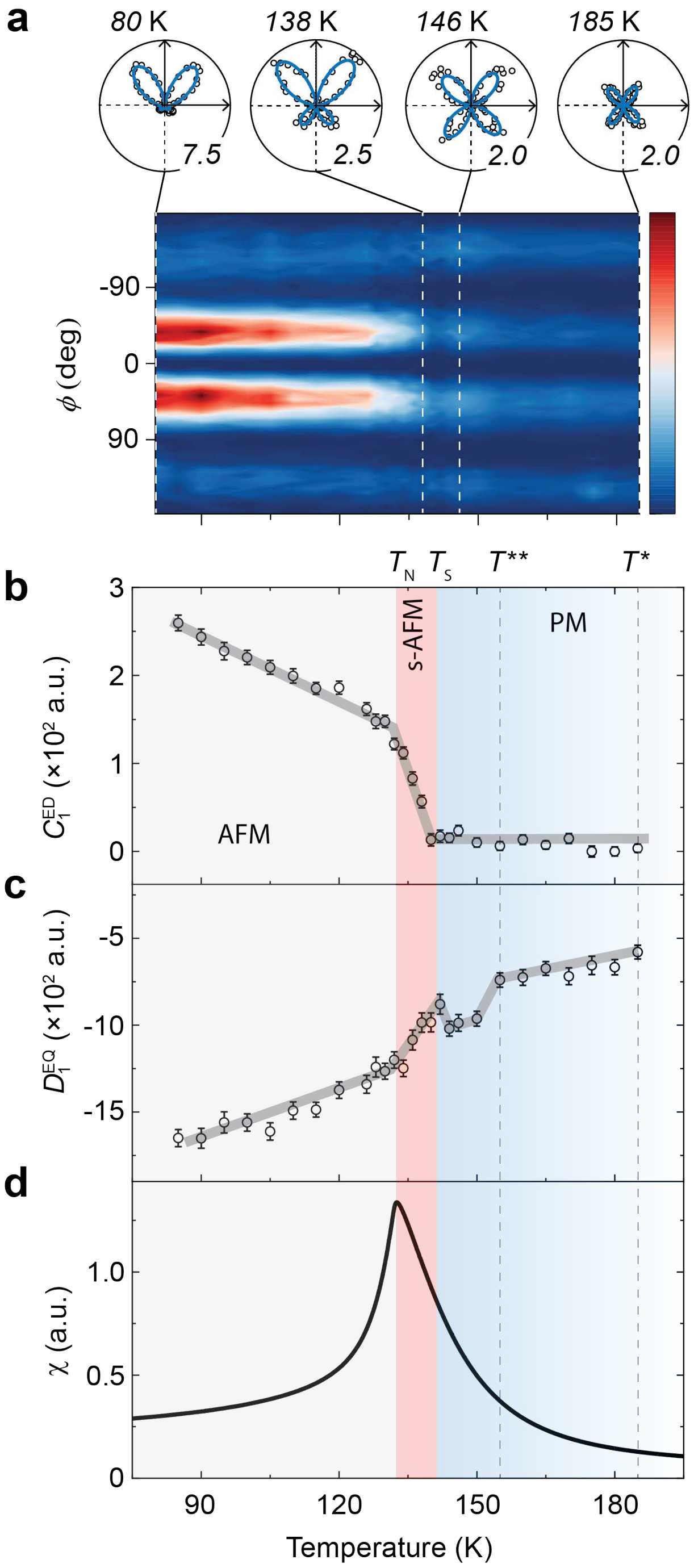

**Figure 6**

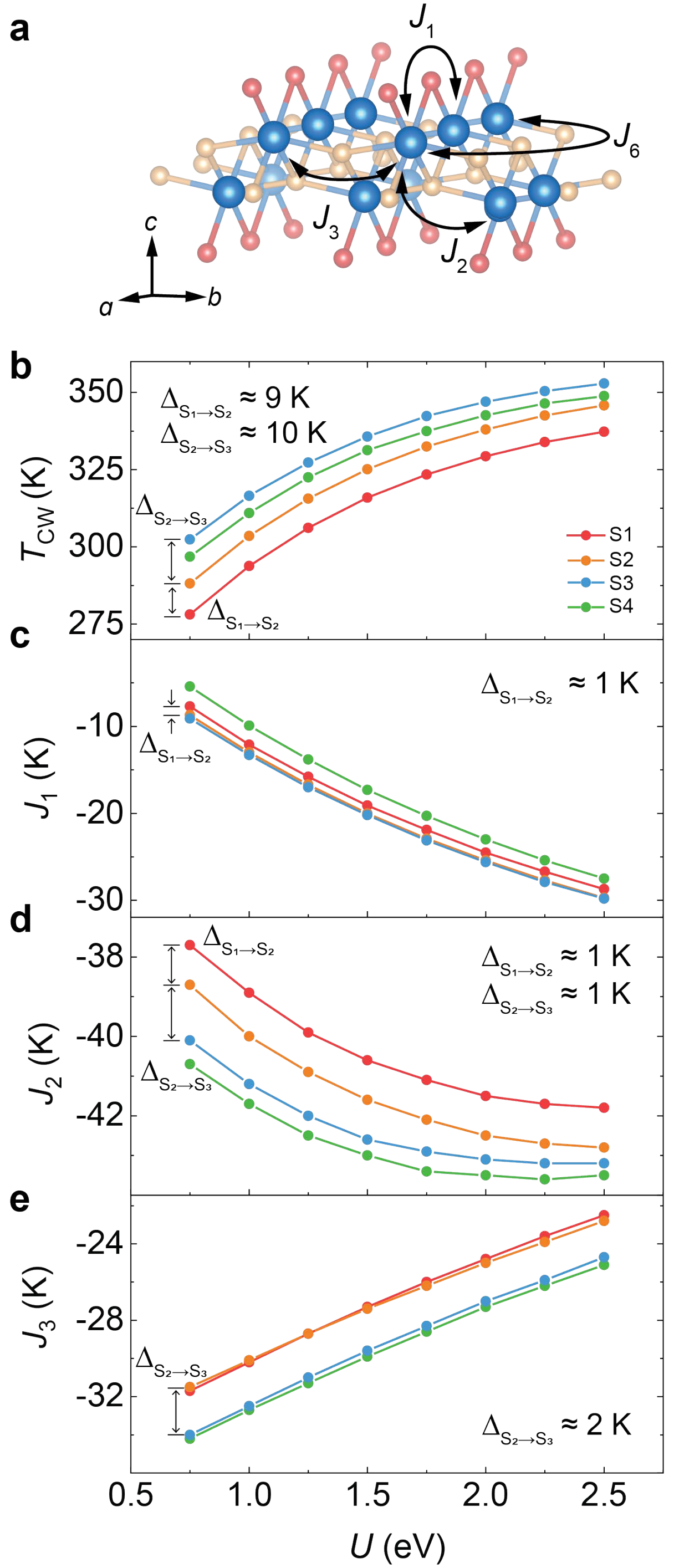

# Supplementary Materials for

## "Extraordinary" Phase Transition Revealed in a van der Waals Antiferromagnet

Xiaoyu Guo[1, Δ] , Wenhao Liu[2, Δ], Jonathan Schwartz[3], Suk Hyun Sung[3], Dechen Zhang[1], Makoto Shimizu[4,*], Aswin L. N. Kondusamy[2], Lu Li[1], Kai Sun[1], Hui Deng[1], Harald O. Jeschke[5], Igor I. Mazin[6], Robert Hovden[3], Bing Lv[2, +] and Liuyan Zhao[1, +]

[1] *Department of Physics, University of Michigan, Ann Arbor, MI 48019, USA*

[2] *Department of Physics, the University of Texas at Dallas, Richardson, TX 75080, USA*

[3] *Department of Materials Science and Engineering, University of Michigan, Ann Arbor, MI 48109, USA*

[4] *Department of Physics, Okayama University, Okayama 700-8530, Japan*

[5] *Research Institute for Interdisciplinary Science, Okayama University, Okayama 700-8530, Japan*

[6] *Department of Physics and Astronomy, and Quantum Science and Engineering Center, George Mason University, Fairfax, VA 22030, USA*

[+] Corresponding to: blv@utdallas.edu, lyzhao@umich.edu

[*] Present address: *Department of Physics, Graduate School of Science, Kyoto University, Kyoto 606-8502, Japan*

[Δ] Authors contribute equally

**Section 1: magnetic susceptibility measurement of bulk CrSBr**

**Section 2: high temperature oblique SHG RA from bulk CrSBr and SHG radiation source determination**

**Section 3: domain survey on bulk CrSBr**

**Section 4: superposition of surface ED and bulk EQ**

**Section 5: temperature dependence of $C_2^{ED}$ and $D_2^{ED}$**

**Section 6: supplementary results from density functional theory calculation**

## Section 1: magnetic susceptibility measurement of bulk CrSBr

We performed magnetic susceptibility measurement on the same sample where the second harmonic generation rotational anisotropy (SHG RA) measurement was performed. As is shown in Figure S1a, apart from the diverging behavior at $T_N$ that indicates the bulk antiferromagnetic (AFM) phase transition, an anomaly is evident at $T_F$ = 30 K, marking the onset of the possible ferromagnetic phase transition. Here, only a weak signature has been observed in our high-quality crystal, consistent with the proposal that this phase transition is related to the magnetic defects inside the crystal. We have also fitted the high temperature (>150 K) magnetic susceptibility using the Curie-Weiss Law:

$$\chi = \chi_0 + \frac{C}{T - T_0},$$

where $\chi_0$ is the temperature-independent susceptibility arising from the background, $C$ is a constant and $T_0$ is the Curie-Weiss temperature (Fig. S1b). The fitted $T_0$ = 152 K.

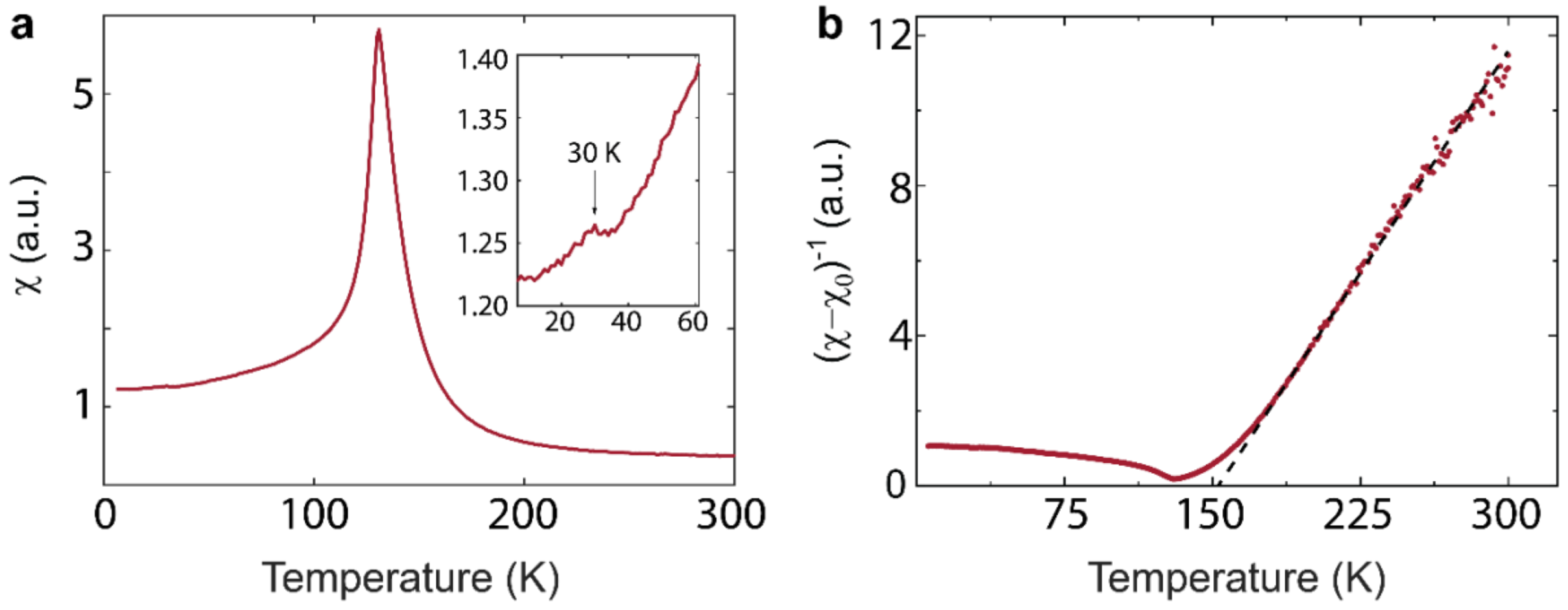

**Figure S1 | Magnetic susceptibility measurement of bulk CrSBr. a**, Magnetic susceptibility measured as a function of temperature. A magnetic field of 1000 Oe was applied along the crystallographic *b*-axis for the measurement. The inset shows the zoom-in region illustrating the anomaly at $T_F$ = 30 K, where the possible ferromagnetic phase transition happens. **b**, Temperature dependence of the inverse magnetic susceptibility. The black dash line shows the fitting of the data using the Curie−Weiss law.

## Section 2: high temperature oblique SHG RA from bulk CrSBr and SHG radiation source determination

Figure S2 shows the SHG RA patterns measured at 293 K and 185 K on the same sample but at different locations. Both sets of the patterns show the same symmetries: two-fold rotational symmetry about the *c*-axis ($C_{2c}$), and mirror symmetries with respect to mirrors perpendicular to *a*-axis ($m_a$) and *b*-axis ($m_b$), consistent with the crystallography point group *mmm*. They also show similar shapes and SHG intensities. The characteristic temperature scale $T^*$ = 185 K that indicates the presence of spin-spin interaction cannot be captured by our SHG RA technique.

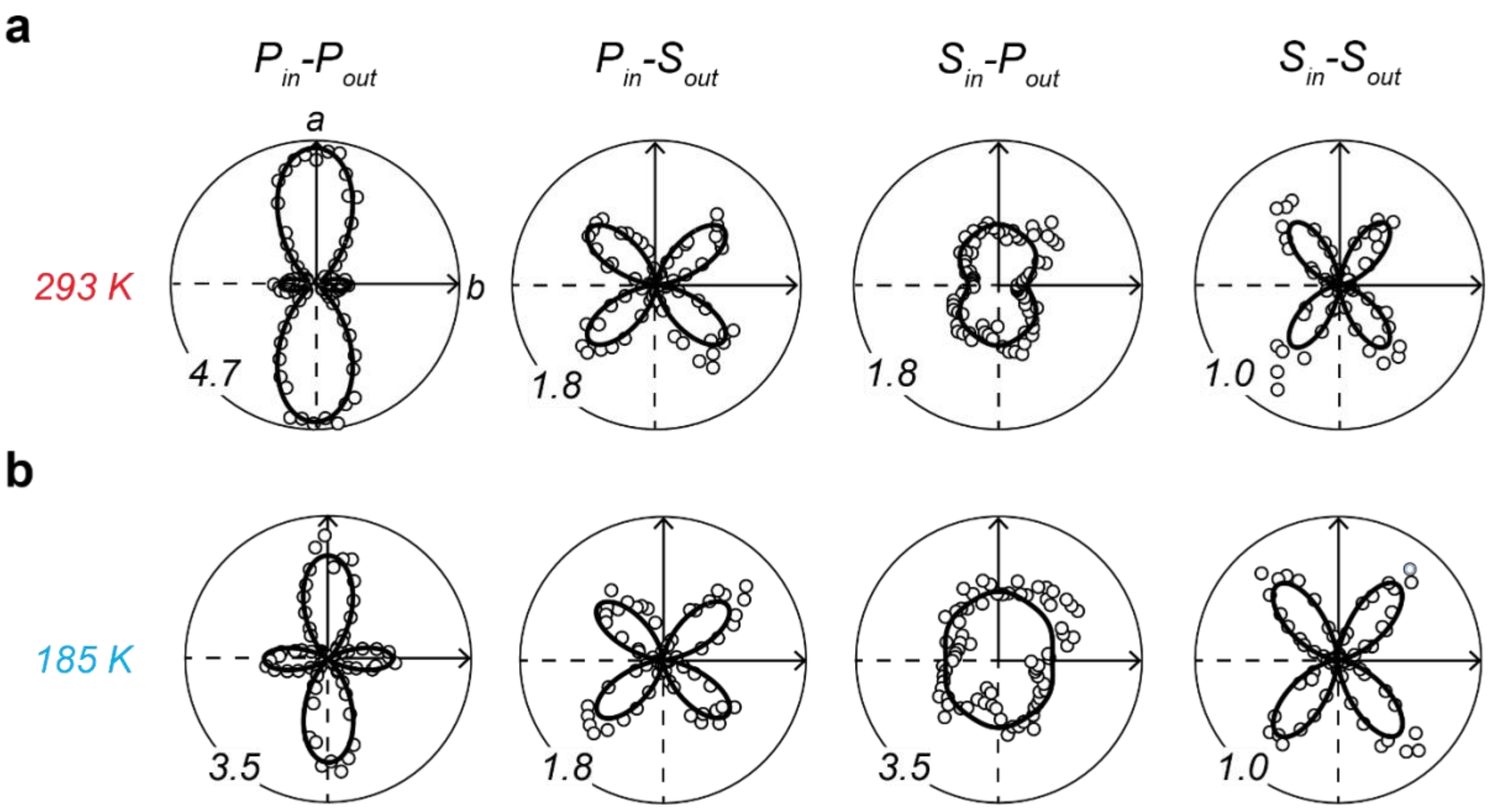

**Figure S2 | SHG RA patterns at high temperatures.** Four channels of SHG RA patterns measured at **a**, 293 K and **b**, 185 K. Experiment data (circles) are fitted by functional forms simulated based on group theory analysis (solid curves). Numbers at the corners indicate the scales of the polar plots.

The experimental data has been fitted with the functional forms simulated from the electric quadrupole (EQ) contribution under the point group *mmm* and shown as solid curves in Figure S2. Other radiation sources including surface electric dipole (ED), bulk magnetic dipole (MD) and electric field induced second harmonic (EFISH) have been ruled out. Figure S3 shows the SHG RA raw data measured at $T$ = 185 K, together with the simulated pattens under bulk EQ (point group *mmm*), surface ED (point group *mm2*), bulk MD (point group *mmm*), and EFISH at the surface, with the induced dipole along the *c*-axis (point group *mmm*), using the functional forms provided. We see that the raw data matches the EQ simulation the best. Specifically, in the other three cases, there is always one channel showing no SHG signal, in contrast with our raw data, where it is present in all four channels. Consequently, we have pinned down bulk EQ as the primary source for our SHG signal.

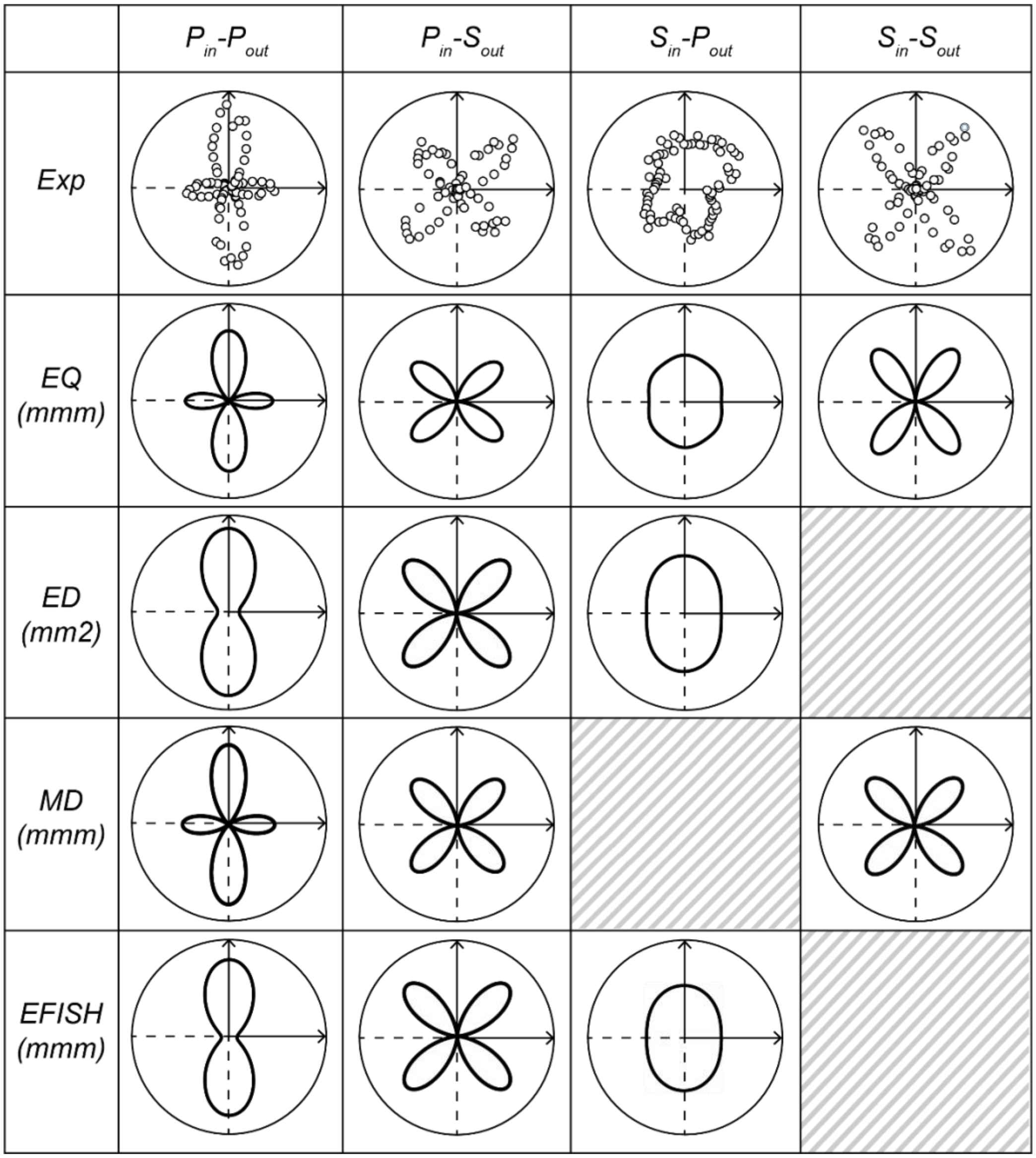

**Figure S3 | Simulation results for various SHG radiation sources.** SHG RA raw data measured at 185 K in all the four channels, together with the simulated pattens from bulk EQ, surface ED, bulk MD and EFISH at the surface.

Here, we provide the simulated functional forms of the SHG RA patterns at the high temperature from different radiation sources under the corresponding point groups that are used to construct Figure S3.

1. **Bulk EQ** under the point group ***mmm***:

The rank-4 nonlinear optical susceptibility tensor has the form:

$$\chi_{mmm}^{EQ} = \begin{pmatrix} \begin{pmatrix} \chi_{xxxx} & 0 & 0 \\ 0 & \chi_{xxyy} & 0 \\ 0 & 0 & \chi_{xxzz} \end{pmatrix} & \begin{pmatrix} 0 & \chi_{xyxy} & 0 \\ \chi_{xxyy} & 0 & 0 \\ 0 & 0 & 0 \end{pmatrix} & \begin{pmatrix} 0 & 0 & \chi_{xzxz} \\ 0 & 0 & 0 \\ \chi_{xxzz} & 0 & 0 \end{pmatrix} \\ \begin{pmatrix} 0 & \chi_{yyxx} & 0 \\ \chi_{yxyx} & 0 & 0 \\ 0 & 0 & 0 \end{pmatrix} & \begin{pmatrix} \chi_{yyxx} & 0 & 0 \\ 0 & \chi_{yyyy} & 0 \\ 0 & 0 & \chi_{yyzz} \end{pmatrix} & \begin{pmatrix} 0 & 0 & 0 \\ 0 & 0 & \chi_{yzyz} \\ 0 & \chi_{yyzz} & 0 \end{pmatrix} \\ \begin{pmatrix} 0 & 0 & \chi_{zzxx} \\ 0 & 0 & 0 \\ \chi_{zxzx} & 0 & 0 \end{pmatrix} & \begin{pmatrix} 0 & 0 & 0 \\ 0 & 0 & \chi_{zzyy} \\ 0 & \chi_{zyzy} & 0 \end{pmatrix} & \begin{pmatrix} \chi_{zzxx} & 0 & 0 \\ 0 & \chi_{zzyy} & 0 \\ 0 & 0 & \chi_{zzzz} \end{pmatrix} \end{pmatrix},$$

leading to the following functional forms for the radiation.

In the $P_{\text{in}}$-$P_{\text{out}}$ channel:

$$\begin{aligned} S_{EQ,mmm}^{PP} = {} & \text{Sin}[\theta]^2(-\chi_{zzzz}\text{Cos}[\theta]\text{Sin}[\theta]^2 - \text{Cos}[\theta]^3(\chi_{zxzx}\text{Cos}[\phi]^2 + \chi_{zyzy}\text{Sin}[\phi]^2) \\ & + 2\text{Cos}[\theta]\text{Sin}[\theta]^2(\chi_{zzxx}\text{Cos}[\phi]^2 + \chi_{zzyy}\text{Sin}[\phi]^2))^2 \\ & + \text{Cos}[\theta]^2(2\text{Cos}[\theta]^2\text{Sin}[\theta](\chi_{xxzz}\text{Cos}[\phi]^2 + \chi_{yyzz}\text{Sin}[\phi]^2) \\ & - \text{Sin}[\theta]^3(\chi_{xzxz}\text{Cos}[\phi]^2 + \chi_{yzyz}\text{Sin}[\phi]^2) - \text{Cos}[\theta]^2\text{Sin}[\theta](\chi_{xxxx}\text{Cos}[\phi]^4 \\ & + (2\chi_{xxyy} + \chi_{xyxy} + \chi_{yxyx} + 2\chi_{yyxx})\text{Cos}[\phi]^2\text{Sin}[\phi]^2 + \chi_{yyyy}\text{Sin}[\phi]^4))^2. \end{aligned}$$

In the $P_{\text{in}}$-$S_{\text{out}}$ channel:

$$\begin{aligned} S_{EQ,mmm}^{PS} = {} & (2(\chi_{xxzz} - \chi_{yyzz})\text{Cos}[\theta]^2\text{Cos}[\phi]\text{Sin}[\theta]\text{Sin}[\phi] - (\chi_{xzxz} \\ & - \chi_{yzyz})\text{Cos}[\phi]\text{Sin}[\theta]^3\text{Sin}[\phi] - \text{Cos}[\theta]^2\text{Sin}[\theta]((\chi_{xxxx} - \chi_{yxyx} \\ & - 2\chi_{yyxx})\text{Cos}[\phi]^3\text{Sin}[\phi] + (2\chi_{xxyy} + \chi_{xyxy} - \chi_{yyyy})\text{Cos}[\phi]\text{Sin}[\phi]^3))^2. \end{aligned}$$

In the $S_{\text{in}}$-$P_{\text{out}}$ channel:

$$\begin{aligned} S_{EQ,mmm}^{SP} = {} & \text{Cos}[\theta]^2\text{Sin}[\theta]^2(\chi_{zyzy}\text{Cos}[\phi]^2 + \chi_{zxzx}\text{Sin}[\phi]^2)^2 \\ & + \text{Cos}[\theta]^2\text{Sin}[\theta]^2(\chi_{xyxy}\text{Cos}[\phi]^4 + (\chi_{xxxx} - 2(\chi_{xxyy} + \chi_{yyxx}) \\ & + \chi_{yyyy})\text{Cos}[\phi]^2\text{Sin}[\phi]^2 + \chi_{yxyx}\text{Sin}[\phi]^4)^2. \end{aligned}$$

In the $S_{\text{in}}$-$S_{\text{out}}$ channel:

$$\begin{aligned} S_{EQ,mmm}^{SS} = {} & \text{Sin}[\theta]^2((\chi_{xyxy} + 2\chi_{yyxx} - \chi_{yyyy})\text{Cos}[\phi]^3\text{Sin}[\phi] + (\chi_{xxxx} - 2\chi_{xxyy} \\ & - \chi_{yxyx})\text{Cos}[\phi]\text{Sin}[\phi]^3)^2, \end{aligned}$$

where $\theta$ is the incident polar angle and $\phi$ the azimuth angle between the scattering plane and the crystallographic $a$-axis.

2. **Surface ED** under the point group $\boldsymbol{mm2}$ and ***i*-type surface ED** under the magnetic point group $\boldsymbol{m'm2'}$:

   The rank-3 nonlinear optical susceptibility tensor has the form:

$$\chi_{mm2}^{ED} = \begin{pmatrix} \begin{pmatrix} 0 \\ 0 \\ \chi_{\mathrm{xxz}} \end{pmatrix} & \begin{pmatrix} 0 \\ 0 \\ 0 \end{pmatrix} & \begin{pmatrix} \chi_{\mathrm{xxz}} \\ 0 \\ 0 \end{pmatrix} \\ \begin{pmatrix} 0 \\ 0 \\ 0 \end{pmatrix} & \begin{pmatrix} 0 \\ 0 \\ \chi_{\mathrm{yyz}} \end{pmatrix} & \begin{pmatrix} 0 \\ \chi_{\mathrm{yyz}} \\ 0 \end{pmatrix} \\ \begin{pmatrix} \chi_{\mathrm{zxx}} \\ 0 \\ 0 \end{pmatrix} & \begin{pmatrix} 0 \\ \chi_{\mathrm{zyy}} \\ 0 \end{pmatrix} & \begin{pmatrix} 0 \\ 0 \\ \chi_{\mathrm{zzz}} \end{pmatrix} \end{pmatrix},$$

   leading to the following functional forms for the radiation.

   In the $P_{\mathrm{in}}$-$P_{\mathrm{out}}$ channel:

$$S_{ED,mm2}^{PP} = 4\mathrm{Cos}[\theta]^4\mathrm{Sin}[\theta]^2\left(\chi_{\mathrm{xxz}}\mathrm{Cos}[\phi]^2 + \chi_{\mathrm{yyz}}\mathrm{Sin}[\phi]^2\right)^2 + \mathrm{Sin}[\theta]^2\left(\chi_{\mathrm{zzz}}\mathrm{Sin}[\theta]^2 + \mathrm{Cos}[\theta]^2\left(\chi_{\mathrm{zxx}}\mathrm{Cos}[\phi]^2 + \chi_{\mathrm{zyy}}\mathrm{Sin}[\phi]^2\right)\right)^2.$$

   In the $P_{\mathrm{in}}$-$S_{\mathrm{out}}$ channel:

$$S_{ED,mm2}^{PS} = 4\left(\chi_{\mathrm{xxz}} - \chi_{\mathrm{yyz}}\right)^2\mathrm{Cos}[\theta]^2\mathrm{Cos}[\phi]^2\mathrm{Sin}[\theta]^2\mathrm{Sin}[\phi]^2.$$

   In the $S_{\mathrm{in}}$-$P_{\mathrm{out}}$ channel:

$$S_{ED,mm2}^{SP} = \mathrm{Sin}[\theta]^2(\chi_{\mathrm{zyy}}\mathrm{Cos}[\phi]^2 + \chi_{\mathrm{zxx}}\mathrm{Sin}[\phi]^2)^2.$$

   In the $S_{\mathrm{in}}$-$S_{\mathrm{out}}$ channel:

$$S_{ED,mm2}^{SS} = 0,$$

3. **Bulk MD** under the point group $\boldsymbol{mmm}$:

   The rank-3 nonlinear optical susceptibility tensor has the form:

$$\chi_{mmm}^{MD} = \begin{pmatrix} \begin{pmatrix}0\\0\\0\end{pmatrix} & \begin{pmatrix}0\\0\\\chi_{xyz}\end{pmatrix} & \begin{pmatrix}0\\\chi_{xyz}\\0\end{pmatrix} \\ \begin{pmatrix}0\\0\\\chi_{yxz}\end{pmatrix} & \begin{pmatrix}0\\0\\0\end{pmatrix} & \begin{pmatrix}\chi_{yxz}\\0\\0\end{pmatrix} \\ \begin{pmatrix}0\\\chi_{zxy}\\0\end{pmatrix} & \begin{pmatrix}\chi_{zxy}\\0\\0\end{pmatrix} & \begin{pmatrix}0\\0\\0\end{pmatrix} \end{pmatrix},$$

leading to the following functional forms for the radiation.

In the $P_{\text{in}}$-$P_{\text{out}}$ channel:

$$S_{MD,mmm}^{PP} = 4\text{Cos}[\theta]^2\text{Sin}[\theta]^2(\text{Cos}[\theta]^4 + \text{Sin}[\theta]^4)(\chi_{yxz}\text{Cos}[\phi]^2 - \chi_{xyz}\text{Sin}[\phi]^2)^2.$$

In the $P_{\text{in}}$-$S_{\text{out}}$ channel:

$$S_{MD,mmm}^{PS} = (\chi_{xyz} + \chi_{yxz} - \chi_{zxy})^2\text{Cos}[\theta]^4\text{Sin}[\theta]^2\text{Sin}[2\phi]^2.$$

In the $S_{\text{in}}$-$P_{\text{out}}$ channel:

$$S_{MD,mmm}^{SP} = 0.$$

In the $S_{\text{in}}$-$S_{\text{out}}$ channel:

$$S_{MD,mmm}^{SS} = \chi_{zxy}^2\text{Sin}[\theta]^2\text{Sin}[2\phi]^2.$$

4. **EFISH** with induced electric dipole along the *c*-axis under the point group ***mmm***:

The rank-4 nonlinear optical susceptibility tensor has the form:

$$\chi_{mmm}^{EFISH} = \begin{pmatrix} \begin{pmatrix}\chi_{xxxx}&0&0\\0&\chi_{xxyy}&0\\0&0&\chi_{xxzz}\end{pmatrix} & \begin{pmatrix}0&\chi_{xyxy}&0\\\chi_{xxyy}&0&0\\0&0&0\end{pmatrix} & \begin{pmatrix}0&0&\chi_{xzxz}\\0&0&0\\\chi_{xxzz}&0&0\end{pmatrix} \\ \begin{pmatrix}0&\chi_{yyxx}&0\\\chi_{yxyx}&0&0\\0&0&0\end{pmatrix} & \begin{pmatrix}\chi_{yyxx}&0&0\\0&\chi_{yyyy}&0\\0&0&\chi_{yyzz}\end{pmatrix} & \begin{pmatrix}0&0&0\\0&0&\chi_{yzyz}\\0&\chi_{yyzz}&0\end{pmatrix} \\ \begin{pmatrix}0&0&\chi_{zzxx}\\0&0&0\\\chi_{zxzx}&0&0\end{pmatrix} & \begin{pmatrix}0&0&0\\0&0&\chi_{zzyy}\\0&\chi_{zyzy}&0\end{pmatrix} & \begin{pmatrix}\chi_{zzxx}&0&0\\0&\chi_{zzyy}&0\\0&0&\chi_{zzzz}\end{pmatrix} \end{pmatrix},$$

leading to the following functional forms for the radiation.

In the $P_{\text{in}}$-$P_{\text{out}}$ channel:

$$\begin{aligned} S_{EFISH,mmm}^{PP} = {} & 4\text{Cos}[\theta]^4\text{Sin}[\theta]^2(\chi_{xzxz}\text{Cos}[\phi]^2 + \chi_{yzyz}\text{Sin}[\phi]^2)^2 \\ & + \text{Sin}[\theta]^2(\chi_{zzzz}\text{Sin}[\theta]^2 + \text{Cos}[\theta]^2(\chi_{zxxz}\text{Cos}[\phi]^2 + \chi_{zyyz}\text{Sin}[\phi]^2))^2 \end{aligned}$$

In the $P_{\text{in}}$-$S_{\text{out}}$ channel:

$$S_{EFISH,mmm}^{PS} = 4(\chi_{\text{xzxz}} - \chi_{\text{yzyz}})^2 \text{Cos}[\theta]^2 \text{Cos}[\phi]^2 \text{Sin}[\theta]^2 \text{Sin}[\phi]^2$$

In the $S_{\text{in}}$-$P_{\text{out}}$ channel:

$$S_{EFISH,mmm}^{SP} = \text{Sin}[\theta]^2 (\chi_{\text{zyyz}} \text{Cos}[\phi]^2 + \chi_{\text{zxxz}} \text{Sin}[\phi]^2)^2$$

In the $S_{\text{in}}$-$S_{\text{out}}$ channel:

$$S_{EFISH,mmm}^{SS} = 0.$$

## Section 3: domain survey on bulk CrSBr

We have surveyed several locations on two pieces of bulk CrSBr samples. Figures S4a and S4b present the optical image of the two CrSBr samples. Figure S4c shows the SHG RA patterns measured in the $P_{in}$-$P_{out}$ channel at the locations numbered and labeled in Figures S4a and S4b. It can be noted that each of the CrSBr sample is a single domain.

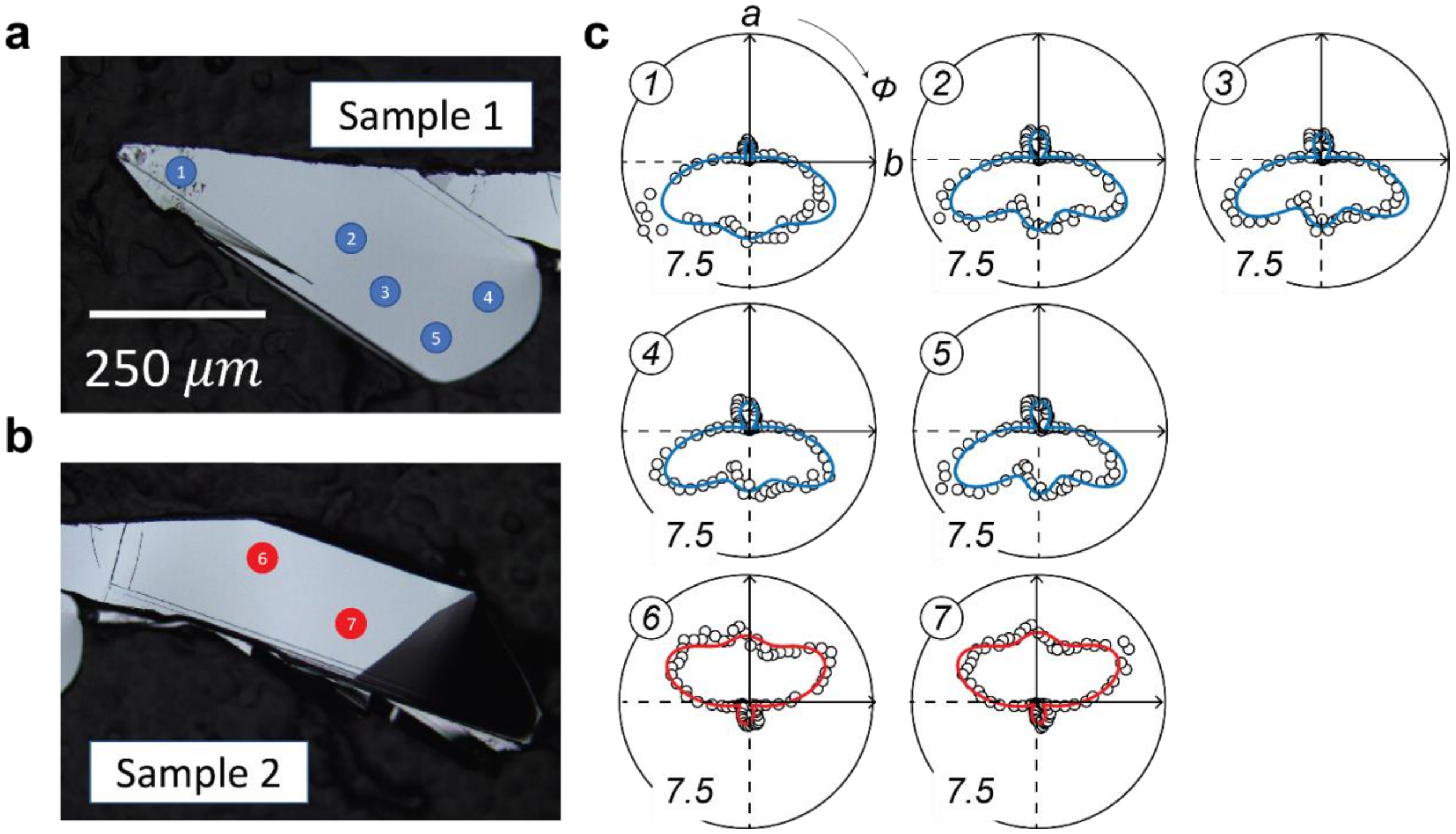

**Figure S4 | Spatial survey of magnetic domains of bulk CrSBr.** Optical image of **a**, sample 1 and **b**, sample 2. **c**, SHG RA in the $P_{in}$-$P_{out}$ channel measured at the locations numbered and labelled in **a** and **b**. The numbers at the bottom indicate the scales of the polar plots.

We also surveyed the SHG RA patterns at the same location on the sample through multiple thermal cycles. Figure S5a shows the SHG RA patterns in the four polarization channels observed through the first cool down. After heating up to 185 K (Fig. S5b) and cool down to 80 K again, a different set of SHG RA patterns are observed (Fig. S5c). The patterns shown in Figures S5a and S5c are related by $m_a$ and $C_{2c}$, which are the relation between the degenerate magnetic domains. This indicates that different magnetic domains are randomly selected through each thermal cycle. We have performed four thermal cycles, one of which shows the flip of the SHG RA patterns.

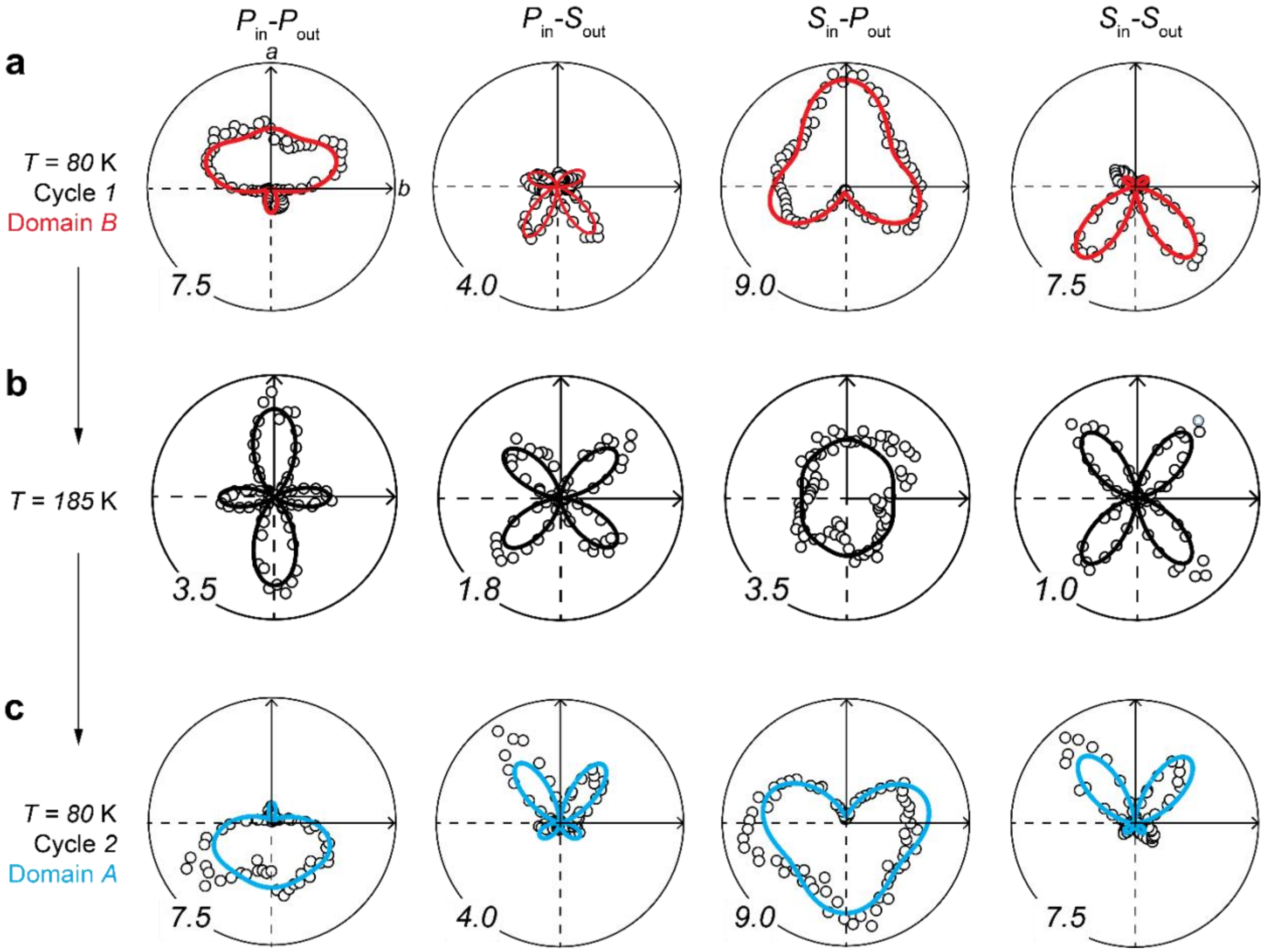

**Figure S5 | Magnetic domains of bulk CrSBr under different thermal cycles.** SHG RA patterns measured at the same location on the sample through multiple thermal cycles. Two sets of patterns, related by $m_a$ and $C_{2c}$, have been observed at 80 K, which come from **a**, domain B and **c**, domain A. **b**, SHG RA patterns measured at $T$ = 185 K. The numbers at the bottom indicate the scales of the polar plots.

**Section 4: superposition of surface ED and bulk EQ**

Here, we provide the functional forms of SHG radiation under the superposition of bulk EQ and surface ED. The nonlinear optical susceptibility tensor for bulk EQ has already been given in Section 2. We now need to consider the time-variant (*c*-type) SHG radiation from the surface under the magnetic point group $\boldsymbol{m'm2'}$:

The rank-3 nonlinear optical susceptibility tensor has the form:

$$\chi_{m'm2'}^{ED,c,A} = \begin{pmatrix} \begin{pmatrix} \chi_{xxx} \\ 0 \\ 0 \end{pmatrix} & \begin{pmatrix} 0 \\ \chi_{xyy} \\ 0 \end{pmatrix} & \begin{pmatrix} 0 \\ 0 \\ \chi_{xzz} \end{pmatrix} \\ \begin{pmatrix} 0 \\ \chi_{yxy} \\ 0 \end{pmatrix} & \begin{pmatrix} \chi_{yxy} \\ 0 \\ 0 \end{pmatrix} & \begin{pmatrix} 0 \\ 0 \\ 0 \end{pmatrix} \\ \begin{pmatrix} 0 \\ 0 \\ \chi_{zxz} \end{pmatrix} & \begin{pmatrix} 0 \\ 0 \\ 0 \end{pmatrix} & \begin{pmatrix} \chi_{zxz} \\ 0 \\ 0 \end{pmatrix} \end{pmatrix}$$

for domain A, and

$$\chi_{m'm2'}^{ED,c,B} = \begin{pmatrix} \begin{pmatrix} -\chi_{xxx} \\ 0 \\ 0 \end{pmatrix} & \begin{pmatrix} 0 \\ -\chi_{xyy} \\ 0 \end{pmatrix} & \begin{pmatrix} 0 \\ 0 \\ -\chi_{xzz} \end{pmatrix} \\ \begin{pmatrix} 0 \\ -\chi_{yxy} \\ 0 \end{pmatrix} & \begin{pmatrix} -\chi_{yxy} \\ 0 \\ 0 \end{pmatrix} & \begin{pmatrix} 0 \\ 0 \\ 0 \end{pmatrix} \\ \begin{pmatrix} 0 \\ 0 \\ -\chi_{zxz} \end{pmatrix} & \begin{pmatrix} 0 \\ 0 \\ 0 \end{pmatrix} & \begin{pmatrix} -\chi_{zxz} \\ 0 \\ 0 \end{pmatrix} \end{pmatrix}$$

for domain B. Note that the rank-3 nonlinear optical susceptibility tensors for domain A and B are related by a minus sign because of the time-reversal relation, leading to the different interference patterns shown in Figure 4 of the main text. The radiation solely from the ***c*-type surface ED** under the magnetic point group $\boldsymbol{m'm2'}$ is:

In the $P_{\text{in}}$-$P_{\text{out}}$ channel:

$$S_{ED,m'm2'}^{PP} = \Big(-2\chi_{zxz}\mathrm{Cos}[\theta]\mathrm{Cos}[\phi]\mathrm{Sin}[\theta]^2 + \mathrm{Cos}[\theta]\Big(\chi_{xzz}\mathrm{Cos}[\phi]\mathrm{Sin}[\theta]^2 + \mathrm{Cos}[\theta]^2\big(\chi_{xxx}\mathrm{Cos}[\phi]^3 + (\chi_{xyy} + 2\chi_{yxy})\mathrm{Cos}[\phi]\mathrm{Sin}[\phi]^2\big)\Big)\Big)^2.$$

In the $P_{\text{in}}$-$S_{\text{out}}$ channel:

$$S_{ED,m'm2'}^{PS} = \Big(\chi_{xzz}\mathrm{Sin}[\theta]^2\mathrm{Sin}[\phi] + \mathrm{Cos}[\theta]^2\Big((\chi_{xxx} - 2\chi_{yxy})\mathrm{Cos}[\phi]^2\mathrm{Sin}[\phi] + \chi_{xyy}\mathrm{Sin}[\phi]^3\Big)\Big)^2.$$

In the $S_{\text{in}}$-$P_{\text{out}}$ channel:

$$S^{SP}_{ED,m'm2'} = \left(\text{Cos}[\theta]\left(\chi_{\text{xyy}}\text{Cos}[\phi]^3 + \left(\chi_{\text{xxx}} - 2\chi_{\text{yxy}}\right)\text{Cos}[\phi]\text{Sin}[\phi]^2\right)\right)^2.$$

In the $S_{\text{in}}$-$S_{\text{out}}$ channel:

$$S^{SS}_{ED,m'm2'} = \left(\left(\chi_{\text{xyy}} + 2\chi_{\text{yxy}}\right)\text{Cos}[\phi]^2\text{Sin}[\phi] + \chi_{\text{xxx}}\text{Sin}[\phi]^3\right)^2.$$

Note that the SHG radiation from domain A and B share the same form. Only the interference between the surface magnetism and the bulk EQ radiations will lead to distinct patterns between domain A and domain B, as is shown below:

Considering the interference between surface **ED** with surface magnetism under the magnetic point group $\boldsymbol{m'm2'}$ and **EQ** under the point group $\boldsymbol{mmm}$:

In the $P_{\text{in}}$-$P_{\text{out}}$ channel:

$$\begin{aligned} S^{PP}_{ED+EQ} = {} & (\text{Sin}[\theta](-2\chi_{\text{zxz}}\text{Cos}[\theta]\text{Cos}[\phi]\text{Sin}[\theta] - \chi_{\text{zzzz}}\text{Cos}[\theta]\text{Sin}[\theta]^2 - \text{Cos}[\theta]^3(\chi_{\text{zxzx}}\text{Cos}[\phi]^2 \\ & + \chi_{\text{zyzy}}\text{Sin}[\phi]^2) + 2\text{Cos}[\theta]\text{Sin}[\theta]^2(\chi_{\text{zzxx}}\text{Cos}[\phi]^2 + \chi_{\text{zzyy}}\text{Sin}[\phi]^2)) \\ & + \text{Cos}[\theta](\chi_{\text{xzz}}\text{Cos}[\phi]\text{Sin}[\theta]^2 + 2\text{Cos}[\theta]^2\text{Sin}[\theta](\chi_{\text{xxzz}}\text{Cos}[\phi]^2 + \chi_{\text{yyzz}}\text{Sin}[\phi]^2) \\ & - \text{Sin}[\theta]^3(\chi_{\text{xzxz}}\text{Cos}[\phi]^2 + \chi_{\text{yzyz}}\text{Sin}[\phi]^2) + \text{Cos}[\theta]^2(\chi_{\text{xxx}}\text{Cos}[\phi]^3 + (\chi_{\text{xyy}} \\ & + 2\chi_{\text{yxy}})\text{Cos}[\phi]\text{Sin}[\phi]^2) - \text{Cos}[\theta]^2\text{Sin}[\theta](\chi_{\text{xxxx}}\text{Cos}[\phi]^4 + (2\chi_{\text{xxyy}} + \chi_{\text{xyxy}} \\ & + \chi_{\text{yxyx}} + 2\chi_{\text{yyxx}})\text{Cos}[\phi]^2\text{Sin}[\phi]^2 + \chi_{\text{yyyy}}\text{Sin}[\phi]^4)))^2. \end{aligned}$$

In the $P_{\text{in}}$-$S_{\text{out}}$ channel:

$$\begin{aligned} S^{PS}_{ED+EQ} = {} & (2(\chi_{\text{xxzz}} - \chi_{\text{yyzz}})\text{Cos}[\theta]^2\text{Cos}[\phi]\text{Sin}[\theta]\text{Sin}[\phi] + \chi_{\text{xzz}}\text{Sin}[\theta]^2\text{Sin}[\phi] - (\chi_{\text{xzxz}} \\ & - \chi_{\text{yzyz}})\text{Cos}[\phi]\text{Sin}[\theta]^3\text{Sin}[\phi] + \text{Cos}[\theta]^2((\chi_{\text{xxx}} - 2\chi_{\text{yxy}})\text{Cos}[\phi]^2\text{Sin}[\phi] \\ & + \chi_{\text{xyy}}\text{Sin}[\phi]^3) - \text{Cos}[\theta]^2\text{Sin}[\theta]((\chi_{\text{xxxx}} - \chi_{\text{yxyx}} - 2\chi_{\text{yyxx}})\text{Cos}[\phi]^3\text{Sin}[\phi] \\ & + (2\chi_{\text{xxyy}} + \chi_{\text{xyxy}} - \chi_{\text{yyyy}})\text{Cos}[\phi]\text{Sin}[\phi]^3))^2. \end{aligned}$$

In the $S_{\text{in}}$-$P_{\text{out}}$ channel:

$$\begin{aligned} S^{SP}_{ED+EQ} = {} & (-\text{Cos}[\theta]\text{Sin}[\theta](\chi_{\text{zyzy}}\text{Cos}[\phi]^2 + \chi_{\text{zxzx}}\text{Sin}[\phi]^2) + \text{Cos}[\theta](\chi_{\text{xyy}}\text{Cos}[\phi]^3 + (\chi_{\text{xxx}} \\ & - 2\chi_{\text{yxy}})\text{Cos}[\phi]\text{Sin}[\phi]^2 - \text{Sin}[\theta](\chi_{\text{xyxy}}\text{Cos}[\phi]^4 + (\chi_{\text{xxxx}} - 2(\chi_{\text{xxyy}} + \chi_{\text{yyxx}}) \\ & + \chi_{\text{yyyy}})\text{Cos}[\phi]^2\text{Sin}[\phi]^2 + \chi_{\text{yxyx}}\text{Sin}[\phi]^4)))^2. \end{aligned}$$

In the $S_{\text{in}}$-$S_{\text{out}}$ channel:

$$\begin{aligned} S^{SS}_{ED+EQ} = {} & ((\chi_{\text{xyy}} + 2\chi_{\text{yxy}})\text{Cos}[\phi]^2\text{Sin}[\phi] + \chi_{\text{xxx}}\text{Sin}[\phi]^3 - \text{Sin}[\theta]((\chi_{\text{xyxy}} + 2\chi_{\text{yyxx}} \\ & - \chi_{\text{yyyy}})\text{Cos}[\phi]^3\text{Sin}[\phi] + (\chi_{\text{xxxx}} - 2\chi_{\text{xxyy}} - \chi_{\text{yxyx}})\text{Cos}[\phi]\text{Sin}[\phi]^3))^2 \end{aligned}$$

for domain A. Domain B shares the similar functional forms with an additional minus sign before all the rank-3 tensor elements $\chi_{ijk}$.

## Section 5: temperature dependence of $C_2^{ED}$ and $D_2^{ED}$

The $C_2^{ED} = \chi_{xxx}$ and $D_2^{EQ} = \chi_{xyxy} + 2\chi_{yyxx} - \chi_{yyyy}$ fitted from the temperature-dependent SHG RA in the $S_{in}$-$S_{out}$ channel are plotted in Figure S6. Both $C_2^{ED}$ and $D_2^{EQ}$ have a relatively large uncertainty. Despite this, $D_2^{EQ}$ is capable of tracking $T^{**}$ and $T_S$, similar as $D_1^{EQ}$ in the Figure 5c. However, unlike $D_1^{EQ}$, $D_2^{EQ}$ cannot capture $T_N$ due to the larger uncertainty.

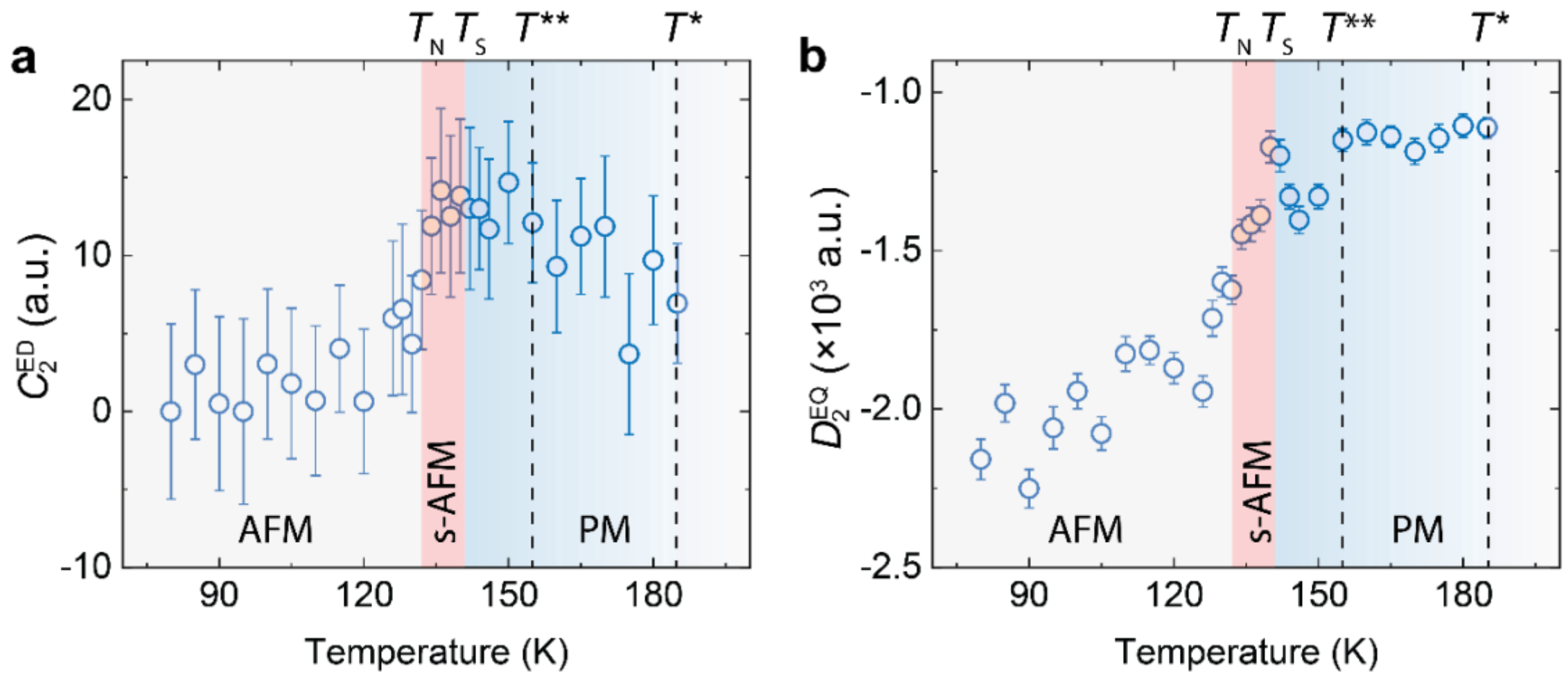

**Figure S6 | Temperature dependence of $C_2^{ED}$ and $D_2^{EQ}$.** **a**, $C_2^{ED}$ and **b**, $D_2^{EQ}$ as a function of temperature fitted from the $S_{in}$-$S_{out}$ channel. The regions of paramagnetism (PM), surface antiferromagnetism (s-AFM) and bulk antiferromagnetism (AFM) are shaded in different colors, with their characteristic temperatures marked.

## Section 6: supplementary results from density functional theory (DFT) calculation

Figure S7 shows the $U$-dependence of $J_6$, which hardly changes from bulk to monolayer CrSBr under various settings.

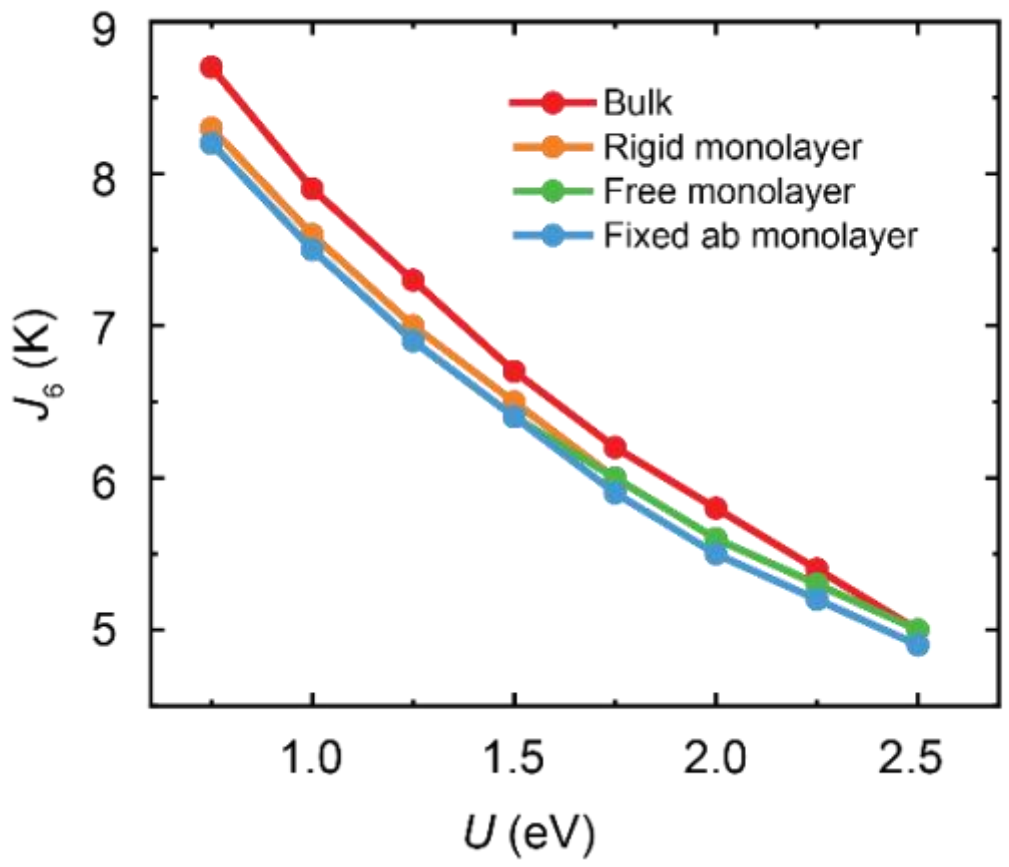

**Figure S7 |** $U$-dependence of $J_6$ for bulk CrSBr (red), rigid monolayer (orange), fixed *ab* monolayer (blue) and free monolayer (green).

Table 1 provides the information from $J_1$ to $J_7$ and the Curie-Weiss temperatures in bulk CrSBr and monolayer CrSBr under various settings. The corresponding Cr-Cr distance for each $J$ has also been provided.

TABLE I. Energy mapping results for CrSBr bulk and monolayers.

bulk

| $U$ (eV) | $J_1$ | $J_2$ | $J_3$ | $J_4$ | $J_5$ | $J_6$ | $J_7$ | $\vartheta_{CW}$ |
|---|---|---|---|---|---|---|---|---|
| 0.75 | -7.7(8) | -37.7(7) | -31.7(3) | -0.4(7) | 0.2(7) | 8.7(4) | -2.4(2) | 278.1 |
| 1.0 | -12.1(7) | -38.9(6) | -30.2(3) | -0.4(6) | 0.1(6) | 7.9(4) | -2.3(2) | 293.8 |
| 1.25 | -15.8(6) | -39.9(5) | -28.7(3) | -0.5(5) | 0.1(5) | 7.3(3) | -2.3(2) | 306.1 |
| 1.5 | -19.1(6) | -40.6(5) | -27.3(2) | -0.5(5) | 0.1(5) | 6.7(3) | -2.3(2) | 315.9 |
| 1.75 | -21.9(5) | -41.1(4) | -26.0(2) | -0.5(4) | 0.1(4) | 6.2(3) | -2.2(2) | 323.4 |
| 2.0 | -24.5(5) | -41.5(4) | -24.8(2) | -0.5(4) | 0.1(4) | 5.8(2) | -2.2(2) | 329.3 |
| 2.25 | -26.7(4) | -41.7(4) | -23.6(2) | -0.5(4) | 0.1(4) | 5.4(2) | -2.1(1) | 333.9 |
| 2.5 | -28.7(4) | -41.8(3) | -22.5(2) | -0.5(3) | 0.1(3) | 5.0(2) | -2.1(1) | 337.3 |
| $d_{Cr-Cr}$ (Å) | 3.544 | 3.592 | 4.738 | 5.917 | 6.166 | 7.089 | 7.603 | |

rigid monolayer

| $U$ (eV) | $J_1$ | $J_2$ | $J_3$ | $J_4$ | $J_5$ | $J_6$ | $J_7$ | $\vartheta_{CW}$ |
|---|---|---|---|---|---|---|---|---|
| 0.75 | -8.7(8) | -38.7(7) | -31.5(3) | -0.5(7) | -0.1(7) | 8.3(4) | -2.4(2) | 288.1 |
| 1.0 | -13.0(7) | -40.0(6) | -30.1(3) | -0.6(6) | -0.1(6) | 7.6(4) | -2.3(2) | 303.5 |
| 1.25 | -16.7(6) | -40.9(5) | -28.7(3) | -0.6(5) | -0.1(5) | 7.0(3) | -2.3(2) | 315.6 |
| 1.5 | -20.0(6) | -41.6(5) | -27.4(2) | -0.6(5) | -0.1(5) | 6.5(3) | -2.2(2) | 325.1 |
| 1.75 | -22.9(5) | -42.1(4) | -26.2(2) | -0.6(4) | -0.1(4) | 6.0(3) | -2.2(2) | 332.5 |
| 2.0 | -25.4(5) | -42.5(4) | -25.0(2) | -0.6(4) | 0.0(4) | 5.6(2) | -2.2(2) | 338.0 |
| 2.25 | -27.7(4) | -42.7(4) | -23.9(29 | -0.6(4) | 0.0(4) | 5.3(2) | -2.1(1) | 342.5 |
| 2.5 | -29.7(4) | -42.8(3) | -22.8(2) | -0.5(3) | 0.0(3) | 5.0(2) | -2.1(1) | 345.8 |
| $d_{Cr-Cr}$ (Å) | 3.544 | 3.592 | 4.738 | 5.917 | 6.166 | 7.089 | 7.603 | |

free monolayer

| $U$ (eV) | $J_1$ | $J_2$ | $J_3$ | $J_4$ | $J_5$ | $J_6$ | $J_7$ | $\vartheta_{CW}$ |
|---|---|---|---|---|---|---|---|---|
| 0.75 | -5.4(8) | -40.7(7) | -34.2(3) | -0.8(7) | -0.2(7) | 8.2(4) | -2.0(2) | 296.8 |
| 1.0 | -9.9(7) | -41.7(6) | -32.7(3) | -0.8(6) | -0.1(6) | 7.5(4) | -1.9(2) | 310.9 |
| 1.25 | -13.8(6) | -42.5(5) | -31.3(3) | -0.9(5) | -0.1(5) | 6.9(3) | -1.9(2) | 322.5 |
| 1.5 | -17.3(6) | -43.0(5) | -29.9(2) | -0.8(5) | -0.1(5) | 6.4(3) | -1.9(2) | 331.3 |
| 1.75 | -20.3(5) | -43.4(4) | -28.6(2) | -0.8(4) | 0.0(4) | 6.0(3) | -1.9(2) | 337.5 |
| 2.0 | -23.0(5) | -43.5(4) | -27.3(2) | -0.7(4) | 0.0(4) | 5.6(2) | -1.9(2) | 342.6 |
| 2.25 | -25.4(4) | -43.6(4) | -26.2(2) | -0.7(4) | 0.1(4) | 5.3(2) | -1.9(1) | 346.4 |
| 2.5 | -27.5(4) | -43.5(3) | -25.1(2) | -0.7(3) | 0.1(3) | 5.0(2) | -1.9(1) | 348.8 |
| $d_{Cr-Cr}$ (Å) | 3.533 | 3.597 | 4.727 | 5.901 | 6.156 | 7.066 | 7.591 | |

fixed $ab$ monolayer

| $U$ (eV) | $J_1$ | $J_2$ | $J_3$ | $J_4$ | $J_5$ | $J_6$ | $J_7$ | $\vartheta_{CW}$ |
|---|---|---|---|---|---|---|---|---|
| 0.75 | -9.1(8) | -40.1(7) | -34.0(3) | -0.7(7) | -0.1(7) | 8.2(4) | -2.1(2) | 302.4 |
| 1.0 | -13.3(7) | -41.2(6) | -32.5(3) | -0.7(6) | -0.1(6) | 7.5(4) | -2.0(2) | 316.5 |
| 1.25 | -17.0(6) | -42.0(5) | -31.0(3) | -0.8(5) | -0.1(5) | 6.9(3) | -2.0(2) | 327.3 |
| 1.5 | -20.2(6) | -42.6(5) | -29.6(2) | -0.7(5) | -0.1(5) | 6.4(3) | -2.0(2) | 335.7 |
| 1.75 | -23.1(5) | -42.9(4) | -28.3(2) | -0.7(4) | -0.1(4) | 5.9(3) | -2.0(2) | 342.3 |
| 2.0 | -25.6(5) | -43.1(4) | -27.0(2) | -0.7(4) | 0.0(4) | 5.5(2) | -2.0(2) | 347.0 |
| 2.25 | -27.9(4) | -43.2(4) | -25.9(2) | -0.7(4) | 0.0(4) | 5.2(2) | -2.0(1) | 350.4 |
| 2.5 | -29.8(4) | -43.2(3) | -24.7(2) | -0.6(3) | 0.1(3) | 4.9(2) | -1.9(1) | 352.9 |
| $d_{Cr-Cr}$ (Å) | 3.544 | 3.599 | 4.738 | 5.917 | 6.171 | 7.089 | 7.606 | |

Table 2 provides the information of the change in the interatomic distances and bond angles in bulk CrSBr and monolayer CrSBr under different settings.

TABLE II. Geometrical parameters for the three most important exchange paths of CrSBr.

| $J_1$ | bulk | fixed $ab$ monolayer | free monolayer | rigid monolayer |
|---|---|---|---|---|
| Cr-Cr distance (Å) | 3.54428 | 3.54428 (±0%) | 3.53296 (-0.32%) | 3.54428 (±0%) |
| Cr-S distance (Å) | 2.39300 | 2.39200 (-0.04%) | 2.39051 (-0.10%) | 2.39295 (-0.00%) |
| Cr-S-Cr angle (°) | 95.5568 | 95.6096 (+0.06%) | 95.2852 ( -0.28%) | 95.5596 (-0.00%) |
| Cr-Br distance (Å) | 2.51795 | 2.52227 (+0.17%) | 2.52066 (+0.11%) | 2.51800 (+0.00%) |
| Cr-Br-Cr angle (°) | 89.4660 | 89.2714 (-0.22%) | 88.9827 (-0.54%) | 89.4634 (-0.00%) |
| $J_2$ | bulk | fixed $ab$ monolayer | free monolayer | rigid monolayer |
| Cr-Cr distance (Å) | 3.59191 | 3.59947 (+0.21%) | 3.59704 (+0.14%) | 3.59182 (-0.00%) |
| Cr-S distance (Å) | 2.40750 | 2.41018 (+0.11%) | 2.40521 (-0.10%) | 2.40749 (-0.00%) |
| Cr-S-Cr angle (°) | 96.8750 | 97.1018 (+0.23%) | 97.1894 (+0.32%) | 96.8736 (-0.00%) |
| $J_3$ | bulk | fixed $ab$ monolayer | free monolayer | rigid monolayer |
| Cr-Cr distance (Å) | 4.73800 | 4.73800 (±0%) | 4.72671 (-0.24%) | 4.73800 (±0%) |
| Cr-S distance (Å) | 2.40750 | 2.41018 (+0.11%) | 2.40521 (-0.10%) | 2.40749 (-0.00%) |
| Cr-S-Cr angle (°) | 159.4781 | 158.7861 (-0.43%) | 158.5902 (-0.56%) | 159.4818 (+0.00%) |